\documentclass[twocolumn,showpacs,preprintnumbers,superscriptaddress,amsmath,amssymb,floatfix]{revtex4}
\usepackage{graphicx}% Include figure files
\usepackage{dcolumn}% Align table columns on decimal point
\usepackage{bm}% bold math
\usepackage{psfrag}
\usepackage{epsfig}
\usepackage{amsmath}
\usepackage{amssymb}
\usepackage{color}
\usepackage{mathbbol}
\usepackage{subfigure}
\usepackage{epstopdf,txfonts,hyperref}

\newcommand{\nn}{\nonumber}

\newcommand{\ket}[1]{|{#1}\rangle}

\newcommand{\beq}{\begin{equation}}
\newcommand{\eeq}{\end{equation}}

\begin{document}

\title{Quantum phases of incommensurate optical lattices due to cavity backaction}

\author{Hessam Habibian}
\affiliation{Theoretische Physik, Universit\"{a}t des Saarlandes, D-66123 Saarbr\"{u}cken, Germany}
\affiliation{Departament de F\'{i}sica, Universitat Aut\`{o}noma de Barcelona, E-08193 Bellaterra, Spain}

\author{Andr\'{e} Winter}
\affiliation{Theoretische Physik, Universit\"{a}t des Saarlandes, D-66123 Saarbr\"{u}cken, Germany}

\author{Simone Paganelli}
\affiliation{Departament de F\'{i}sica, Universitat Aut\`{o}noma de Barcelona, E-08193 Bellaterra, Spain}
\affiliation{International Institute of Physics, Universidade Federal do Rio Grande do Norte, 59012-970 Natal, Brazil}

\author{Heiko Rieger}
\affiliation{Theoretische Physik, Universit\"{a}t des Saarlandes, D-66123 Saarbr\"{u}cken, Germany}

\author{Giovanna Morigi}
\affiliation{Theoretische Physik, Universit\"{a}t des Saarlandes, D-66123 Saarbr\"{u}cken, Germany}

\date{\today}

\begin{abstract}
Ultracold bosonic atoms are confined by an optical lattice inside an optical resonator and interact with a cavity mode, whose wave length is incommensurate with the spatial periodicity of the confining potential. We predict that the intracavity photon number can be significantly different from zero when the atoms are driven by a transverse laser whose intensity exceeds a threshold value and whose frequency is suitably detuned from the cavity and the atomic transition frequency. In this parameter regime the atoms form clusters in which they emit in phase into the cavity. The clusters are phase locked, thereby maximizing the intracavity photon number. These predictions are based on a Bose-Hubbard model, whose derivation is here reported in detail. The Bose-Hubbard Hamiltonian has coefficients which are due to the cavity field and depend on the atomic density at all lattice sites. The corresponding phase diagram is evaluated using Quantum Monte Carlo simulations in one-dimension and mean-field calculations in two dimensions. Where the intracavity photon number is large, the ground state of the atomic gas lacks superfluidity and possesses finite compressibility, typical of a Bose-glass. 
\end{abstract}

% 03.75.Hh 	Static properties of condensates; thermodynamical, statistical, and structural properties 
%37.30.+i  	Atoms, molecules, and ions in cavities (see also 42.50.Pq Cavity quantum electrodynamics; micromasers)
% 05.30.Jp 	Boson systems
% 32.80.Qk 	Phonons or vibrational states in low-dimensional structures and nanoscale materials
% 42.50.Lc 	Quantum fluctuations, quantum noise, and quantum jumps 


\maketitle

\section{Introduction}

Selforganization of interacting systems of photons and atoms in cavities is a remarkable example of pattern formation in the quantum world. The structures which are formed are due to the nonlinear dependence of the atomic potential on the atomic density: The field scattered into the cavity depends on and determines the atomic density distribution by means of the mechanical effects of atom-photon interactions \cite{Review_RMP}. Features related to these dynamics were already predicted and then observed in atomic gases confined by optical lattices in free space \cite{Deutsch:1995,Hemmerich,Phillips}. In a high-finesse cavity the effect is significantly enhanced due to the strong coupling one can achieve between atoms and light at the single photon level. Here, spatial ordering \cite{Domokos2002,Black2003,Baumann2010,Barrett2012}, collective-atom recoil lasing \cite{CARL:1,CARL:2}, synchronization \cite{Kuramoto}, and motion-induced bistability \cite{Hemmerich:Bistable,Ritter,Stamper-Kurn,Larson:2008} have been observed in gases of laser-cooled atoms inside a high-finesse cavity, when either the cavity or the atoms are driven by an external laser. These phenomena typically occur when the laser intensity exceeds a threshold value and can be revealed by the light transmitted by the resonator's mirrors. 

\begin{figure}
\centering
\includegraphics[width=5cm]{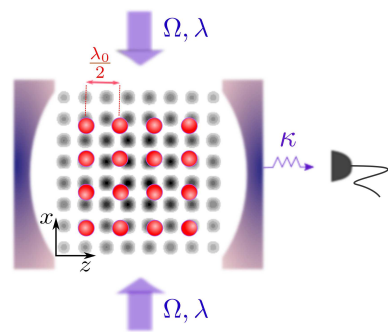}
\caption[]{Ultracold atoms are tightly confined by an optical lattice of periodicity $\lambda_0/2$. They are driven by a weak transverse laser at Rabi frequency $\Omega$ and strongly couple to the mode of a standing-wave cavity both at wavelength $\lambda$. Since $\lambda$ and $\lambda_0$ are incommensurate, one would expect no coherent scattering into the cavity mode.  The mechanical effects due to multiphoton scattering, however, give rise to an incommensurate quantum potential, which mediates an effective long-range interaction between the atoms and modifies the properties of the quantum ground state. As a result, the intracavity photon number can be large. The corresponding  ground state can show features typical of a Bose-glass. } \label{fig:model}
\end{figure}

In recent theoretical studies it was argued that ultracold atoms in cavity quantum electrodynamics setups offer a novel setting to study disorder and glassiness \cite{Lev,Goldbart}. Along a similar line, in a recent work we analysed selforganization of atoms trapped by an external optical lattice and interacting with the standing-wave field of a cavity, whose periodicity is incommensurate with the lattice periodicity. A sketch of the setup is shown in Fig. \ref{fig:model}. In this situation we showed that scattering of light into the resonator generates an optical lattice which is incommensurate with the one confining the atoms, and whose intensity depends on the atomic density at all lattice sites \cite{Hessam:PRL}. This global potential is a feature of cavity quantum electrodynamics setups, where the cavity photons undergo multiple scattering and give rise to an effective atomic potential which is long ranged. It is important to note that this potential would not be generated if the atoms were pointlike particles located at the minima of the classical lattice: in this case there is no coherent scattering into the cavity mode \cite{Habibian2011}. Kinetic energy here favours elastic scattering of photons into the cavity field, giving rise to the formation of patterns which maximize the intracavity field cavity and which can exhibit finite compressibility with no long-range coherence. This latter feature is typical of disordered systems and corresponds to a Bose-glass phase for sufficiently deep potentials \cite{Fisher,BoseGlass,Deissler2010}. Here, it emerges due to the nonlocal quantum potential of the cavity field. 

The purpose of the present paper is to provide the details of  the analytical and numerical derivation at the basis of Ref. \cite{Hessam:PRL}. We discuss in detail the statistical features which emerge from the interaction with the cavity field, which induces an interparticle potential whose range is as large as the system size, and compare our results with numerical studies, in which the quantum ground state properties of bosonic atoms in a classical bichromatic potential are analysed. We also report in detail the experimental parameter regimes where these dynamics can be found making reference to the setup in Ref. \cite{Baumann2010}. 

This paper is organized as follows. In Sec.~\ref{sec:model} the theoretical model is introduced. Here, the Bose-Hubbard Hamiltonian is derived, which includes the infinitely-ranged cavity potential and is at the basis of the numerical simulations. The results for the ground state properties of the Bose-Hubbard model are reported in Sec.~\ref{sec:results}:  In Sec. \ref{sec:results:MC} we show the results of Quantum Monte Carlo simulations for the one-dimensional lattice, while in Sec.~\ref{sec:results:MF} the phase diagrams for the two-dimensional case are discussed, which have been obtained using a mean-field approach. The limits of validity of the model and the experimental parameter are analysed in Sec. \ref{Sec:Validity}. The conclusions are drawn in Sec.~\ref{Sec:Conclusion}, while the appendices report details at the basis of the derivation of the Bose-Hubbard model in Sec. ~\ref{sec:model}  and of the numerical results in Sec. \ref{sec:results:MC}.

\section{Theoretical Model}
\label{sec:model}

The system we consider is composed by $N$ ultracold identical atoms of mass $m$ which obey Bose-Einstein statistics. The atoms are tightly confined by a two-dimensional optical lattice of wave number $k_0=2\pi/\lambda_0$, with $\lambda_0$ the wave length, such that the spatial periodicity is $d_0=\lambda_0/2$. An optical dipole transition of the atoms is driven by a laser and scatters photons into a mode of a high-finesse resonator, according to the geometry shown in Fig.~\ref{fig:model}. The resonator field is a standing wave of wave length $\lambda$ which is incommensurate with the wave length $\lambda_0$ of the external potential confining the atoms. When quantum fluctuations can be neglected, i.e., deep in the Mott-insulator (MI) phase of the external potential, the cavity field is in the vacuum \cite{Habibian2011}. Kinetic energy, on the other hand, induces photon scattering into the cavity field, giving rise to the formation of patterns which maximize scattering into the cavity mode. 

In order to provide an appropriate description we consider the Hamiltonian of the system in second quantization. We derive an effective model for the dynamics of the atomic external degrees of freedom, from which we obtain a Bose-Hubbard Hamiltonian. This Bose-Hubbard Hamiltonian is the starting point of the numerical investigations in Sec. \ref{sec:results}. 

\subsection{Coherent dynamics}

The atoms are prepared in an electronic ground state which we denote by $\ket{1}$. They are confined on the $x-z$ plane by an external potential, and the motion along the $y$ axis is here assumed to be frozen out. For an atom at position ${\bf r}=(x,z)$ the external potential reads
\begin{equation}
\label{V:cl}
V_{\rm cl}({\bf r})=V_0\{\cos^2(k_0z)+\beta\cos^2(k_0x)\}\,,
\end{equation}
where $V_0$ is the potential depth along the $z$ direction and $\beta V_0$ the potential depth along $x$. The atoms are at ultralow temperature $T$ and tightly bound to the potential minima. The quantum gas density also spatially overlaps with the field of an optical resonator: An atomic dipole transition with ground state $\ket{1}$ and excited state $\ket{2}$ at frequency $\omega_0$ couples strongly with a cavity mode at frequency $\omega_c$, wave length $\lambda$, and wave number $k=2\pi/\lambda$ such that the wave vector is along the $z$ axis. The intracavity field is pumped by the photons that the atoms scatter, when these are driven by a transverse laser at frequency $\omega_L$ close to $\omega_c$ such that it has effectively the same wave length $\lambda$ as the cavity mode. The setup is shown in Fig. \ref{fig:model}. 

The coherent dynamics of the cavity field and the atomic internal and external degrees of freedom is governed by Hamiltonian $\hat{\mathcal H}$, which we decompose into the sum of the Hamiltonian for the cavity, the atoms, and their mutual interaction: 
$$\hat{\mathcal H}=\hat{\mathcal H}_C+\hat{\mathcal H}_A+
\hat{\mathcal H}_{\rm int}\,.$$
The Hamiltonian for the cavity mode reads 
\begin{equation}
\label{H:c}
\hat{\mathcal H}_C=\hbar\omega_c\hat{a}^{\dagger}\hat{a}\,,
\end{equation}
where $\hat{a}$ and $\hat{a}^{\dagger}$ are the annihilation and creation operators of a cavity photon, respectively, and obey the commutation relation $[\hat{a},\hat{a}^{\dagger}]=1$. 

The Hamiltonian for the atomic degrees of freedom $\hat{\mathcal H}_A$ (in absence of the resonator) takes the form
\begin{eqnarray}
\hat{\mathcal H}_A
&=&\sum_{j=1,2}\int d^2{\bf r} \,\hat\Psi^\dag_j({\bf r})\hat{H}_j({\bf r})\hat\Psi_j({\bf r})\nonumber\\
&  &+2U_{12}\int d^2{\bf r} \,\hat\Psi^\dag_1({\bf r})\hat\Psi^\dag_2({\bf r})
\hat\Psi_2({\bf r})\hat\Psi_1({\bf r}) \,,
\end{eqnarray}
and is written in terms of the atomic field operator $\hat\Psi_{j}({\bf r},t)$, which destroys an atom in the internal state $\ket{j=1,2}$ at position ${\bf r}$ and time $t$, and obeys the commutation relations $[\hat\Psi_i({\bf r},t),\hat\Psi^\dag_j({\bf r'},t)]=\delta_{ij}\,\delta({\bf r-r'})$. Here, 
\begin{equation}
\hat H_j({\bf r})=-\frac{\hbar^2\nabla^2}{2m}+V_{\rm cl}^{(j)}({\bf r})+\frac{U_{jj}}{2}\hat\Psi^\dag_j({\bf r})\hat\Psi_j({\bf r})+\hbar\omega_0\delta_{j,2}\,,
\end{equation}
where $V_{cl}^{(j)}({\bf r})$ is the optical potential of the atoms in state $j=1,2$, which for the ground state, $j=1$, coincides with $V_{\rm cl}$ in Eq. \eqref{V:cl}, $\delta_{j,2}$ is the Kronecker delta, and $U_{j,l}$ is the strength of the contact interaction between atoms in states $j$ and $l$, with $j,l=1,2$. 

Finally, the Hamiltonian describing the interaction between the atomic dipoles and the electric fields reads
\begin{eqnarray}
\label{Hint}
\hat{\mathcal H}_{\rm int}&=&\hbar g_0\int d^2{\bf r} \cos{(kz)} \left(\hat\Psi^\dag_2({\bf r})\hat\Psi_1({\bf r})\hat a+{\rm H.c.}\right)\nn\\
&&+\hbar \Omega\int d^2{\bf r}\cos(k \,x)\left(\hat\Psi^\dag_2({\bf r})\hat\Psi_1({\bf r}){\rm e}^{-{\rm i}\omega_Lt}+{\rm H.c.}\right)\,,
\end{eqnarray}
where $g_0$ is the cavity vacuum Rabi frequency and the term in the second line describes the coherent coupling between the dipolar transition and a standing-wave laser along the $x$ direction with Rabi frequency $\Omega$. 

\subsection{Heisenberg-Langevin equation and weak excitation limit}

Throughout this paper we assume that the photon scattering processes are elastic. This regime is based on assuming that the detuning between fields and atoms is much larger than the strength with which they are mutually coupled. The large parameter is the detuning 
\begin{equation}
\label{Heiko:1}
\Delta_a=\omega_L-\omega_0
\end{equation}
 between the pump and the atomic transition frequency, which is chosen so that $|\Delta_a|\gg\gamma$, where $\gamma$ the radiative linewidth of the excited state, and so that $|\Delta_a|\gg \Omega,g_0\sqrt{n_{\rm cav}}$, namely, the detuning is much larger than the strength of the coupling between the ground and excited state, where $n_{\rm cav}=\langle \hat{a}^{\dagger}\hat{a}\rangle$ is the intracavity photon number. In this regime the population of the excited state is neglected. 

Photons are elastically scattered into the resonator when the laser is quasi-resonant with the cavity field, which here requires that $|\Delta_a|\gg |\delta_c|$ with 
\begin{equation}
\label{Heiko:2}
\delta_c=\omega_L-\omega_c\,.
\end{equation}  
In this limit the field operator ${\hat\Psi}_2({\bf r},t)$ is a function of the cavity field operator $\hat{a}$ and of the atomic field operator ${\hat\Psi}_1({\bf r},t)$ at the same instant of time according to the relation \cite{Larson2008,Fernandez2010}
\begin{align}\label{Psi_2}
{\hat\Psi}_2({\bf r},t)=&\frac{g_0}{\Delta_a}\cos(kz)\hat\Psi_1({\bf r},t)\, \hat a(t)+\frac{\Omega}{\Delta_a}\cos(k x)\hat\Psi_1({\bf r},t)\,,
\end{align}
which is here given to lowest order in the expansion in $1/|\Delta_a|$. Using Eq. \eqref{Psi_2} in the Heisenberg equation of motion for the field operator $\hat\Psi_1({\bf r},t)$ results in the equation
\begin{align}\label{Psi1_with_a}
\dot{\hat\Psi}_1=&\,-\frac{\rm i}{\hbar}[\hat\Psi_1,\hat{\mathcal H}_A]\nn\\
&-{\rm i}\frac{\Omega^2}{\Delta_a}\cos^2(k x)\hat\Psi_1-{\rm i}U_0\cos^2(kz) \hat a^\dag\hat\Psi_1 \hat a\nn\\
&-{\rm i}S_0\cos(kz) \cos(k x) \Big(\hat a^\dag\hat\Psi_1+\hat\Psi_1 \hat a\Big)\,,
\end{align}
which determines the dynamics of the system together with the Heisenberg-Langevin equation for the cavity field:
\begin{align}\label{a_with_Psi1}
\dot{\hat a}=&-\kappa \hat a+{\rm i}(\delta_c-U_0\hat{\mathcal Y}) \hat a-{\rm i}S_0\hat{\mathcal Z}+\sqrt{2\kappa}\hat{a}_{\rm in}\,,
\end{align}
where $\kappa$ is the cavity linewidth and $\hat{a}_{\rm in}(t)$ is the input noise operator, with $\langle \hat{a}_{\rm in}(t)\rangle=0$ and $\langle \hat{a}_{\rm in}(t)\hat{a}_{\rm in}^\dagger(t')\rangle=\delta(t-t')$ \cite{HLE}. The other parameters are the frequency $U_0=g_0^2/\Delta_a$, which scales the depth of the intracavity potential generated by a single photon,  and the frequency $S_0=g_0\Omega/\Delta_a$, which is the Raman scattering amplitude with which a single photon is scattered by a single atom between the cavity and the laser mode \cite{Maschler}. Moreover, in Eq. \eqref{a_with_Psi1} we have introduced the operators
\begin{align}\label{XYZ_int}
&\hat{\mathcal Z}=\int d^2{\bf r} \cos(kz)\cos\left(kx\right)\, \hat n({\bf r})\,,\nn\\
&\hat{\mathcal Y}=\int d^2{\bf r}\cos^2(kz)\, \hat n({\bf r})\,,
\end{align}
 where 
 \begin{equation}
 \hat n({\bf r})=\hat\Psi_1^\dag({\bf r})\hat\Psi_1({\bf r})
 \end{equation}
is the atomic density. The operators in Eq. \eqref{XYZ_int} count the number of atoms, weighted by the spatial-mode function of the fields and of the corresponding intensity. In the limit in which the atoms can be considered pointlike, then $ \hat n({\bf r})\approx n_{\rm cl}({\bf r})=\sum_j\delta({\bf r}-{\bf r}_j)$ and ${\mathcal Z}_{\rm cl}=\sum_j\cos(kz_j)\cos\left(kx_j\right)$, ${\mathcal Y}_{\rm cl}=\sum_j\cos^2(kz_j)$. Hence, when the atoms are randomly distributed in the cavity field potential then ${\mathcal Z}_{\rm cl}\to 0$. In this case, thus, no photon is elastically scattered into the cavity mode and the cavity field is in the vacuum. This behavior can also be found in the situation we consider in this work, where the atoms are ordered in an array with periodicity which is incommensurate with the periodicity of the pump and cavity standing wave. The focus of this work is to analyze the effect of quantum fluctuations on this behaviour.

%where $\hat n({\bf r})=\hat\Psi_1^\dag({\bf r})\hat\Psi_1({\bf r})$.

\subsection{Adiabatic elimination of the cavity field}

We now derive an effective Hamiltonian governing the motion of the atoms inside the resonator by eliminating the cavity degrees of freedom from the atomic dynamics. This is performed by assuming that the cavity field follows adiabatically the atomic motion. Formally, this consists in a time-scale separation. We identify the time-scale $\Delta t$ over which the atomic motion does not significantly evolve while the cavity field has relaxed to a state which depends on the atomic density at the given interval of time. This requires that 
$|\delta_c+{\rm i}\kappa|\Delta t\gg 1$ while $\kappa_BT\ll \hbar/\Delta t$, with $k_B$ Boltzmann constant \cite{Fernandez2010}. Moreover, the coupling strengths between atoms and fields, which determine the time scale of the evolution due to the mechanical effects of the interaction with the light, are much smaller than $1/\Delta t$. In this limit, we identify the field operator $\hat{a}_{\rm st}$, which is defined by the equation
$$\int_t^{t+\Delta t}\hat{a}(\tau)d\tau/\Delta t\approx \hat{a}_{\rm st}\,,$$
such that  $\int_t^{t+\Delta t}\dot{\hat a}_{\rm st}(\tau)d\tau=0$, with $\dot{\hat a}$ given in Eq. \eqref{a_with_Psi1}. The "stationary" cavity field is a function of the atomic operators at the same (coarse-grained) time, and in particular takes the form
\begin{equation}\label{a_mf}
\hat{a}_{\rm st}=\frac{S_0\hat{\mathcal Z}}{(\delta_c-U_0\hat{\mathcal Y})+{\rm i}\kappa}+\frac{{\rm i}\sqrt{2\kappa}\bar{\hat a}_{\rm in}}{(\delta_c-U_0\hat{\mathcal Y})+{\rm i}\kappa}\,,
%\left[\int_{t_0}^{t_0+\Delta t} a_{in}(\tau){\rm d}\tau\right]/\Delta t\nonumber
\end{equation}
with $\bar{\hat a}_{\rm in}$ the input noise averaged over $\Delta t$. The quantum noise term can be neglected when the mean intracavity photon number is larger than its fluctuations, that corresponds to taking $|S_0\langle\hat{\mathcal{Z}}\rangle|\gg\kappa$. In this limit, the field at the cavity output, 
\begin{equation}
\label{a:out}
\hat{a}_{\rm out}=\sqrt{2\kappa}\hat{a}_{\rm st}-\bar{\hat a}_{\rm in}\,,
\end{equation}
allows one to monitoring the state of the atoms \cite{HLE,MekhovPRL,Habibian2011}. Using Eq. \eqref{a_mf} in place of the field $\hat{a}$ in Eq. \eqref{Psi1_with_a} leads to an equation of motion for the atomic field operator which depends solely on the atomic variables \cite{Larson2008,Fernandez2010}. 

\subsection{Bose-Hubbard model}

Denoting the number of lattice sites by $K$, a well-defined thermodynamic limit is identified assuming the scaling of the cavity parameters with $K$ according to the relations $S_0=s_0/\sqrt{K}$ and $U_0=u_0/K$ \cite{Asboth2005,Fernandez2010}.  Under the assumption that the atoms are tightly bound by the external periodic potential in Eq. \eqref{V:cl}, we apply the single-band approximation and perform the Wannier decomposition of the atomic field operator \cite{Jaksch:BH,Bloch:RMP},
\begin{equation}\label{Psi_expand}
 \hat\Psi_1({\bf r})=\sum_{i,j}w_{i,j}({\bf r}) \hat b_{i,j}\,,
\end{equation}
with the Wannier function $w_{i,j}({\bf r})$ centered at  a lattice site with coordinate $(x_i,z_j)$ (with $x_i=id_0$, $z_j=jd_0$ and $d_0=\lambda_0/2$ the lattice periodicity), while $\hat b_{i,j}$ and $\hat b^\dag_{i,j}$ are the bosonic operators annihilating and creating, respectively, a particle at the corresponding lattice site. The decomposition is performed starting from the equation of motion of the atomic field operator, obtained from Eq. \eqref{Psi1_with_a} with the substitution $\hat{a}\to \hat{a}_{\rm st}$, Eq. \eqref{a_mf}. The details of the procedure are reported in Refs. \cite{Larson2008,Fernandez2010}, and are summarized in Appendix \ref{AppBH}. The resulting Bose-Hubbard Hamiltonian reads \cite{Hessam:PRL}
\begin{eqnarray}\label{Hamil_BH}
\hat{\mathcal H}_{\rm BH}&=&-\sum_{\langle i'j',ij\rangle}\hat{t}_{ij,i'j'}(\hat{b}^\dag_{i,j}\hat{b}_{i',j'}+\hat{b}^\dag_{i',j'}\hat{b}_{i,j})+\frac{U}{2}\sum_{i,j}\hat{n}_{i,j}(\hat{n}_{i,j}-1)\nn\\
&&+\sum_{i,j}\hat{\epsilon}_{i,j} \hat{n}_{i,j}\,
\end{eqnarray}
where the $\langle i'j',ij\rangle$ in the sum denotes the nearest neighbors of the corresponding lattice site. The onsite interaction, the onsite energy, and the tunneling rate are defined as
\begin{eqnarray}
U&=&U_{11}\int d^2{\bf r}\, w_{i,j}({\bf r})^4  \label{U}\,,\\
\hat\epsilon_{i,j}&=& \epsilon^{(0)}+\delta\hat{\epsilon}_{i,j}\,,\nn\\
\hat{t}_{ij,i'j'}&=& t^{(0)}+\delta\hat{t}_{ij,i'j'}\,.
\end{eqnarray}
In particular, the onsite energy (tunneling rate) is the sum of a term which is constant, $\epsilon^{(0)}$ ($t^{(0)}$), and of a term which depends on the lattice site and is due to the cavity field. In detail, the constant terms read
\begin{eqnarray*}
&&\epsilon^{(0)}=\,E_0+V_{0}X_0\,,\\
&&t^{(0)}=-E_1-V_{0}X_1\,,\nonumber
\end{eqnarray*}
with 
\begin{eqnarray}\label{Xst}
&&X_{s}=\int d^2{\bf r}\, w_{i,j}({\bf r})\left[\cos^2(k_0x)+\beta\cos^2(k_0z)\right] w_{i',j'}({\bf r}) \,,\\
&&E_{s}=-\frac{\hbar^2}{2m}\int d^2{\bf r}\, w_{i,j}({\bf r})\nabla^2w_{i',j'}({\bf r})  \label{E}\,,
\end{eqnarray}
such that for $s=0$ then $(i,j)=(i',j')$, while for $s=1$ then $(i',j')$ is a nearest-neighbour site. These terms are due to the dynamics in absence of the cavity field. In appendix \ref{AppBH} we show that the site-dependent term of the tunneling coefficient, $\delta\hat{t}_{ij,i'j'}$ is negligible, so that $\hat{t}_{ij,i'j'}\sim t^{(0)}$. The site-dependent term of the coefficient $\epsilon_{i,j}$ results instead to be relevant and reads
\begin{align}\label{DeltaMu}
\delta\hat{\epsilon}_{i,j}=&\,V_1 J_{0}^{(i,j)}
+\frac{\hbar s_0^2}{\hat\delta_{\rm eff}^2+\kappa^2}\hat\Phi\hat\delta_{\rm eff} Z_{0,0}^{(i,j)}\,,
\end{align}
where the various terms of the sum have different physical origins (note that the term is essentially real under the assumption that $\langle\hat{\delta}_{\rm eff}\rangle\gg\kappa$, which is what we will assume in what follows). The first term on the right-hand side (RHS) is due to the standing wave of the classical transverse pump, with $V_1=\Omega^2/\Delta_a$ and 
\begin{align}\label{paramwann1}
J_{0}^{(i,j)}=\int d^2{\bf r}\, w_{i,j}({\bf r})\cos^2(k \,x)\, w_{i,j}({\bf r})\,.
\end{align} 
This term is site-dependent along the $x$ direction, namely, in the direction of propagation of the transverse field, while it is constant along the $z$ direction when $x$ is fixed. The other terms on the RHS of Eq. \eqref{DeltaMu} are due to the cavity field. In particular, 
\begin{align}\label{paramwann1:2}
&Y_{0}^{(i,j)}=\int d^2{\bf r}\, w_{i,j}({\bf r})\cos^2(kz)\,w_{i,j}({\bf r})\,,\nn\\
&Z_{0,0}^{(i,j)}=\int d^2{\bf r}\, w_{i,j}({\bf r})\cos(kz)\cos(k x)\,w_{i,j}({\bf r})\, ,
\end{align} 
are the overlap integrals due to the cavity optical lattice and the mechanical potential associated with the scattering of cavity photons, respectively, while 
\begin{equation}\label{delta:eff}
\hat\delta_{\rm eff}=\delta_c-u_0 \sum_{i,j}Y_{0}^{(i,j)}\hat{n}_{i,j}/K
\end{equation}
is an operator, whose mean value gives the shift of the cavity resonance due to the atomic distribution \cite{Larson2008}. All these terms are multiplied by the operator 
\begin{equation}\label{Phi}
\hat\Phi=\sum_{i,j} Z_{0,0}^{(i,j)}\hat{n}_{i,j}/K \,,
\end{equation}
which is the sum of the atomic density over the lattice mediated by the Raman scattering amplitude. 

\subsection{Discussion}

The Hamiltonian we have derived reduces, when the pump is off, $\Omega=0$, to the typical Bose-Hubbard model as it occurs in systems of ultracold atoms confined by optical lattices \cite{Jaksch:BH,Bloch:RMP}. The latter exhibits a Superfluid-Mott insulator quantum phase transition which is either controlled by changing the potential depth $V_0$, and hence the hopping coefficient $t$, or the onsite interaction strength $U$ \cite{Jaksch:BH,Bloch:RMP}. In this paper we assume $U$ to be constant and vary $t$ by varying the potential depth $V_0$. 

When the transverse laser drives the cavity field by means of elastic scattering processes, the Hamiltonian depends on the nonlocal operator \eqref{Phi}, which  originates from the long-range interaction between the atoms mediated by the cavity field. The physical observable which is associated with this operator is the cavity field amplitude, 
\begin{equation}\label{a_mf:1}
\hat a_{\rm st}\approx\frac{S_0K\hat{\Phi}}{\hat{\delta}_{\rm eff}+{\rm i}\kappa}\,,
\end{equation}
as is visible by using Eq. \eqref{Phi} in Eq. \eqref{a_mf}, and after discarding the noise term, assuming this is small.
It can be measured by homodyne detection of the field at the cavity output \cite{Homodyne}. The intracavity photon number, $\hat n_{\rm cav}=\hat a_{\rm st}^{\dagger}\hat a_{\rm st}$, reads
\begin{equation}
\label{n:cav}
\hat n_{\rm cav}\approx \frac{S_0^2K^2}{\hat \delta_{\rm eff}^2+\kappa^2}\hat{\Phi}^2\equiv  K\frac{s_0^2}{\hat \delta_{\rm eff}^2+\kappa^2}\hat{\Phi}^2\,,
\end{equation}
and the intensity of the field at the cavity output provides a measurement of operator $\hat {\Phi}^2$, where the second expression on the RHS uses the chosen scaling of the cavity parameters with the number of sites. The intracavity photon number vanishes when the atomic gas forms a MI state: In this case $\langle \hat\Phi^2\rangle_{\rm MI} \propto (\sum_{i,j}Z_0^ {(i,j)})^2=0$, since there is no coherent scattering into the cavity mode. Also deep in the SF phase $\langle \hat\Phi^2\rangle_{\rm SF} \to 0$ as one can verify using Eq. \eqref{Phi}.

It is interesting to note that, using definitions \eqref{Phi} and \eqref{delta:eff}, Hamiltonian \eqref{Hamil_BH} can be cast in the form
\begin{eqnarray}\label{Hamil_BH:1}
\hat{\mathcal H}_{\rm BH}&=&-\sum_{\langle i'j',ij\rangle}t\,(\hat{b}^\dag_{i,j}\hat{b}_{i',j'}+\hat{b}^\dag_{i',j'}\hat{b}_{i,j})+\frac{U}{2}\sum_{i,j}\hat{n}_{i,j}(\hat{n}_{i,j}-1)\nn\\
&&+\epsilon^{(0)} \hat{N}+V_1\sum_{i,j} J_{0}^{(i,j)}\hat{n}_{i,j}+\frac{\hbar s_0^2}{\hat\delta_{\rm eff}^2+\kappa^2}K\hat\Phi^2\hat\delta_{\rm eff}\,,\nn\\
\end{eqnarray}
%{\bf GM: changed the BH due to the cavity -corrected the sign in front of $V_1$.}
where we have neglected the site-dependence of the tunneling parameter (the validity of this approximation is discussed in appendix \ref{AppBH}). In this form the Bose-Hubbard Hamiltonian depends explicitly on the operator corresponding to the number of intracavity photons, see Eq. \eqref{n:cav}, showing the long-range interacting potential due to the cavity field. This potential either decreases or increases the total energy depending on the sign of $\langle \hat\delta_{\rm eff}\rangle$: When $\langle \hat\delta_{\rm eff}\rangle<0$, disordered density distributions are expected when the density is fractional. Its sign hence critically determines whether "disordered" (i.e., aperiodic) density distributions are energetically favourable.  A similar behaviour has been identified in the dynamics of selforganization in periodic potentials \cite{Asboth2005,Nagy2008,Keeling2010}. The dependence of the chemical potential on the operator $\hat\Phi$ is a peculiar property of our model, that makes it differ from  the case of a bichromatic optical lattice \cite{Damski,Deissler2010}, in which the strength of the incommensurate potential is an external parameter, independent of the phase of the ultracold atomic gas. 

We further remark that interesting dynamics could be observed for density distributions such that $\langle\hat{\delta}_{\rm eff}\rangle=0$. In this regime, bistability due to the quantum motion is expected \cite{Larson2008,Ritter,StamperKurn,Stamper-Kurn,Meystre}. In this work we focus on the regime in which the system is far away from this situation, so that $|\langle\hat{\delta}_{\rm eff}\rangle|\gg\kappa$.

\section{Results}
\label{sec:results}

The Bose-Hubbard model of Eq. \eqref{Hamil_BH} is at the basis of the results of this section. We first consider a one-dimensional lattice along the cavity axis by taking the aspect ratio $\beta\gg1$ in $V_{\rm cl}({\bf r})$  and study the phase diagram by means of Monte Carlo simulation. We then analyse the situation where the atoms are ordered in a two-dimensional optical lattice inside the cavity and determine the phase diagram by using a mean-field approach. In  both cases, the phase diagram is found by evaluating the ground state  $|\phi_G\rangle$ of the free-energy, such that it fulfills the relation
\begin{equation}
 \min\left\{\langle\phi_G| \hat{\mathcal{H}}_{\rm BH}-\mu\hat{N} |\phi_G\rangle\right\}\,,
\end{equation}
where $\mu$ is the chemical potential. 

In the following the ratio between the typical interparticle distance $d_0$ and the wave length of the cavity is  chosen to be $d_0/\lambda=83/157$, which is close to $1/2$. Although this ratio is a rational number, nevertheless, for sufficiently small system sizes (here about 300 sites per axis) the emerging dynamics simulates the incommensurate behaviour. We remark that, for the chosen number of sites, the number of intracavity photon is zero for pointlike scatterers when the density is uniform.  We refer the reader to Refs. \cite{Roux08,Deng08}, where the phase diagram of bichromatic potentials in systems of finite size is discussed. 

The parameters of the cavity field, which determine the coefficients of the Bose-Hubbard Hamiltonian in Eq. \eqref{Hamil_BH}, are extracted from the experimental values $g_0/2\pi=14.1$ MHz, $\kappa/2\pi=1.3$ MHz, and  $\gamma/2\pi=3$ MHz for $^{87}$Rb atoms \cite{Baumann2010,Ritter}.  From these values, after fixing the size of the lattice we get $u_0$ and the range of parameters within which we vary the rescaled pump strength $s_0$. Finally, the onsite interaction in the 1D case is $U/\hbar\sim50$ Hz ($U_{11}/\hbar=6.4\times10^{-6}\, \text{Hz\,m}$) and has been taken from Ref.~\cite{Volz_03}. For the 2D optical lattice, $U/\hbar$ varies between 1 and 3 kHz ($U_{11}/\hbar=5.5\times 10^{-11}\, \text{Hz\,m}^2$), see Ref.~\cite{Kruger_07}. A detailed discussion on the validity of Eq. \eqref{Hamil_BH} for this choice of parameters is reported in Sec. \ref{Sec:Validity}

\subsection{One-dimensional lattice}
\label{sec:results:MC}

We focus here on atoms confined in the lowest band of a one-dimensional lattice (1D) along the cavity axis. For this geometry the first term of the RHS of Eq. \eqref{DeltaMu} is a constant energy shift along the cavity axis and can be reabsorbed in the chemical potential, that is $\mu\to \mu-\epsilon^{(0)}-V_1J_0$. The one-dimensional Hamiltonian can be thus written as
\begin{eqnarray}\label{Hamil_BH:1D}
\hat{\mathcal H}_{\rm BH}^{(1D)}&=&-\sum_it\,(\hat{b}^\dag_{i}\hat{b}_{i+1}+\hat{b}^\dag_{i+1}\hat{b}_{i})+\frac{U}{2}\sum_{i}\hat{n}_{i}(\hat{n}_{i}-1)\nn\\
&&+\frac{\hbar s_0^2}{\hat\delta_{\rm eff}^2+\kappa^2}K\hat\Phi^2\hat\delta_{\rm eff}\,,
\end{eqnarray}
%{\bf GM: changed the BH due to the cavity, corrected sign in front of $V_1$.}
where $i$ labels the lattice site along the lattice and $J_0$ is the value of integral \eqref{paramwann1} at the considered string. Here, the onsite energy term depends on the sites only through cavity QED effects. 

\subsubsection{Tunneling coefficient $t\to 0$}

\begin{figure}
\centering
\includegraphics[width=7cm]{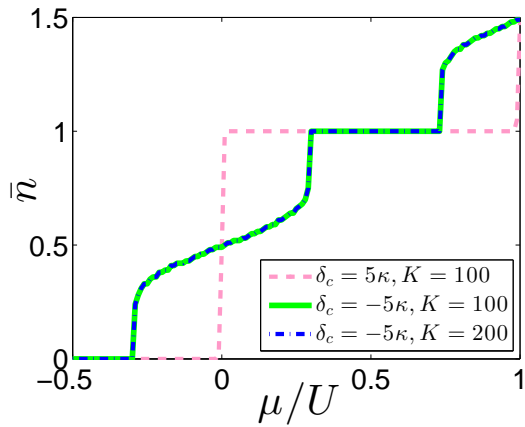}
\caption[]{(color online) Mean density $\bar{n}$ as a function of the chemical $\mu$ (in units of $U$) at $t=0$ in a 1D lattice. The curves have been obtained by an exact diagonalization of Hamiltonian \eqref{Hamil_BH:1D} for  (dashed line) $\delta_c=5\kappa$ and (solid line) $\delta_c=-5\kappa$, both for $K=100$. The other parameters are $s_0=0.006\kappa$ (with $\kappa=2\pi\times 1.3$ MHz), $u_0=0.8\kappa$, and $U/\hbar=50$ Hz.  Here, the chemical potential is reported without the constant shift, i.e., $\mu\to \mu-\epsilon^{(0)}-V_1J_0$. The dash-dotted line has been evaluated for the same parameters of the solid line, except with $K=200$. It shows that the results remain invariant as the system size is scaled up.}\label{Fig:N_mu_Dc}
\end{figure}

We first analyse the case in which the tunneling $t\to 0$, where the atoms are classical pointlike particles localized at the minima of the external potential. We determine the mean density $\bar{n}=\sum_{i=1}^K\langle\hat{n}_i\rangle/K$ as a function of the chemical potential $\mu$ and evaluated over the ground state, which is found by diagonalizing Hamiltonian~\eqref{Hamil_BH:1D} after setting $t=0$ \cite{Footnote:QMC}. The $\mu$-dependency of the density is shown in Fig. \ref{Fig:N_mu_Dc} for different values of the laser-cavity detuning $\delta_c=\omega_L-\omega_c$. The derivative of the curve gives the compressibility $\chi=\partial \bar n/\partial\mu$.
The two curves in Fig. \ref{Fig:N_mu_Dc} correspond to two behaviours that are determined by the sign of the coefficient  $\mathcal C=\langle\hat\delta_{\rm eff}\rangle$ in Eq. \eqref{DeltaMu}. For the parameters we choose this sign is controlled by the sign of the detuning $\delta_c$, Eq. \eqref{Heiko:1}, namely, on whether the laser frequency is tuned to the red or to the blue of the cavity frequency (the parameter choice is discussed in Sec. \ref{Sec:Validity}). 
When $\delta_c>0$, for finite intracavity photon number the cavity-induced interaction energy is positive: The configurations minimizing the energy are thus the ones for which $\langle \hat n_{\rm cav}\rangle=0$, for which Hamiltonian \eqref{Hamil_BH:1D} reduces to the Bose-Hubbard model for atoms in  a periodic potential. 

\begin{figure}
\centering
\includegraphics[width=9cm]{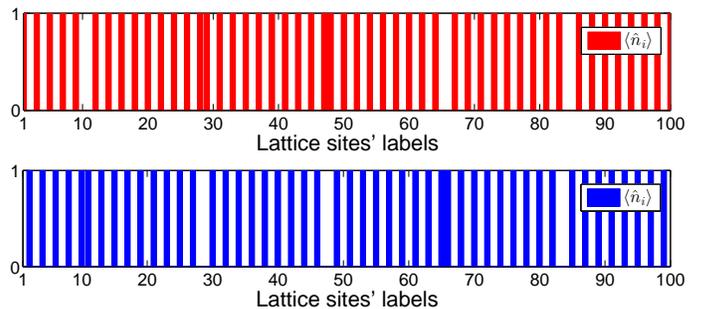}
\caption[]{(color online) Boson occupation number, $n_i$, as a function of the site $i$ for the two distributions corresponding to the two ground states at $\delta_c=-5\kappa$ and $\mu=0$. The other parameters are the same as in Fig. ~\ref{Fig:N_mu_Dc}. The filled stripes correspond to $\langle\hat{n}_i\rangle=1$, the white stripes to $\langle\hat{n}_i\rangle=0$.\label{Fig:Dens_twoConfig}}
\end{figure}

For $\delta_c<0$, on the other hand, the cavity-induced interaction energy is negative:  In this case an arrangement of the atoms that maximizes the intracavity field is energetically favourable. Concomitantly the incompressible phase at $\overline{n}=1$ shrinks, while for fractional densities the compressibility becomes non-zero and the intracavity field is significantly different from zero. For incommensurate densities the quantum ground state is twofold degenerate. Figure \ref{Fig:Dens_twoConfig} displays the two corresponding density profiles (i.e., the local boson occupation numbers $n_i$ as a function of the lattice site index $i$) for the case $\delta_c<0$ and $\mu=0$, for which $\bar n<1$. One configuration corresponds to particle occupation, $n_i=1$, at the sites with $Z_0^{(i,j)}>0$ (hence $\Phi>0$) and $n_i=0$ for the other sites, the other configuration to particle occupation at the sites with $Z_0^{(i,j)}<0$ (hence $\Phi<0$). The two configurations correspond to two phases of the cavity field which differ by $\pi$. This behaviour is analogous to the one encountered in selforganization of ultracold atoms in optical potentials \cite{Black2003,Mottl}. Nevertheless, it must be noticed that whereas in the system of Ref. \cite{Mottl} the atomic patterns are periodic and maximize scattering into the resonator, here scattering into the cavity is maximized by aperiodic density distributions. 

Clearly, even in presence of finite intracavity fields, the mean value of the amplitude vanishes because of this degeneracy. Therefore, when analyzing the intracavity field amplitude we will plot the mean value of operator $|\hat{\Phi}|$. In the rest of this article $\langle \hat{\Phi}\rangle\equiv \langle |\hat{\Phi}|\rangle$ and can be either zero or a positive real number.  

\begin{figure}
\centering
\includegraphics[width=7cm]{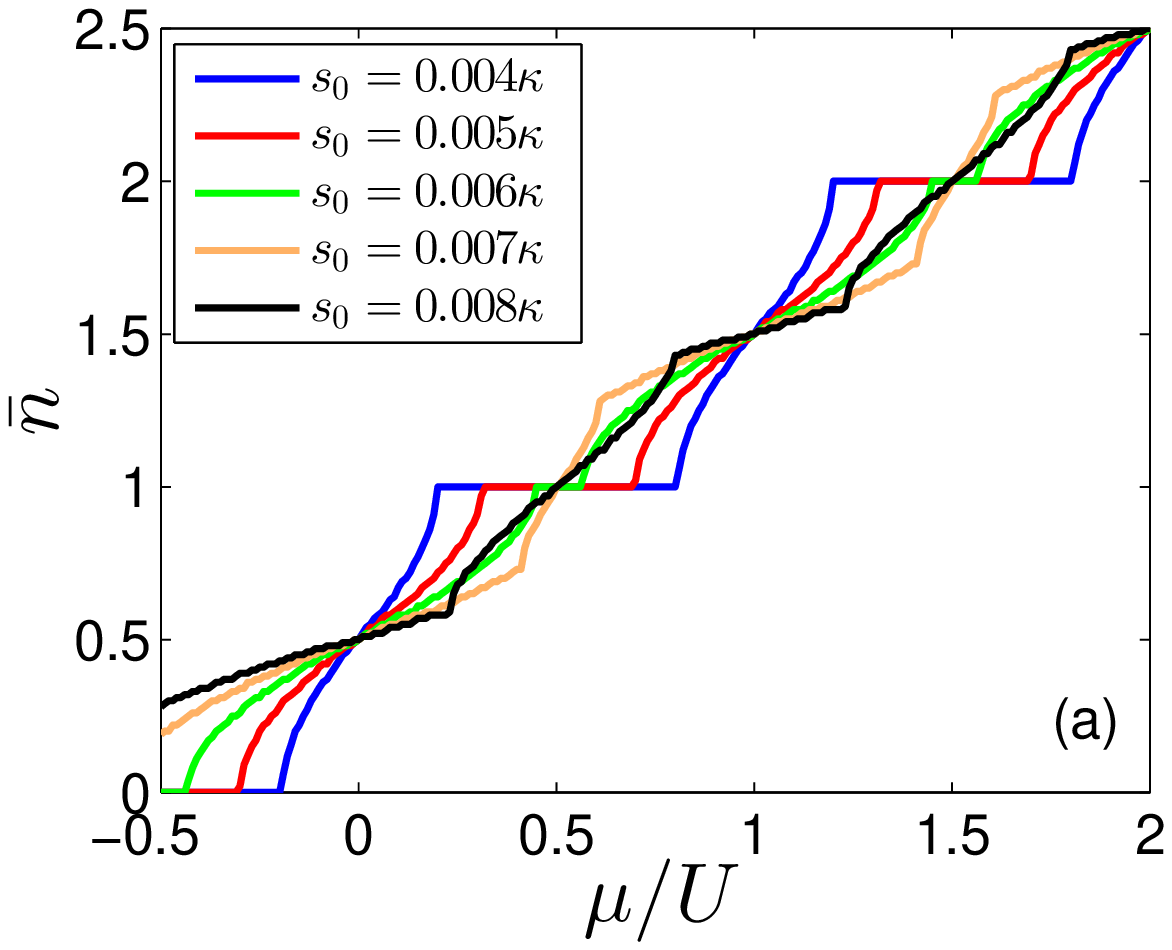}
\includegraphics[width=7cm]{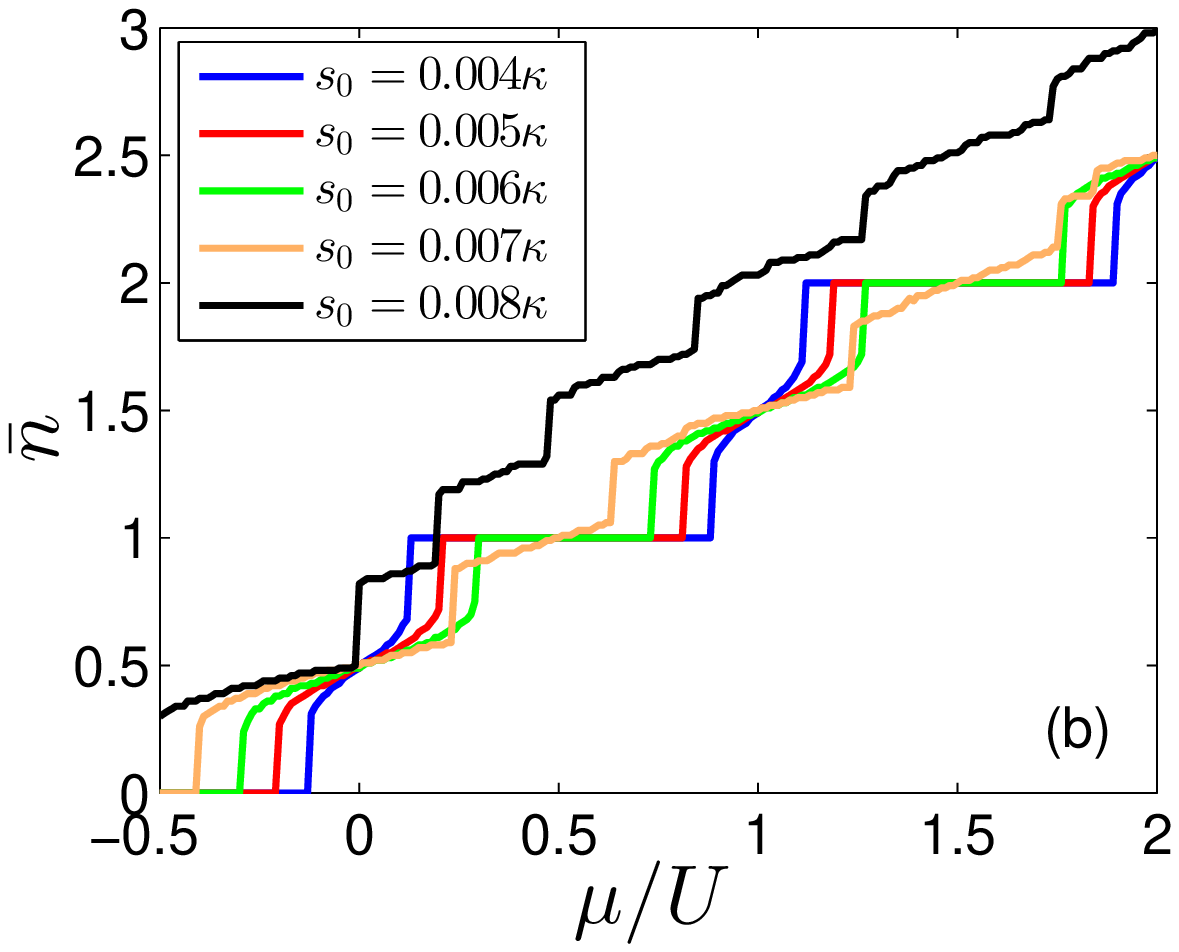}
\caption[]{(color online) Mean density $\bar{n}$ as a function of $\mu$ (in units of $U$) for a 1D lattice of $K=100$ sites for $t=0$, $\delta_c=-5\kappa$, $u_0=0.8\kappa$, $U/\hbar=50$ Hz (with $\kappa/2\pi=1.3$ MHz), while the values of $s_0$ are reported in the legend. The curves in (a) are evaluated by diagonalizing Eq. \eqref{Hamil_BH:1D} after setting $\langle\hat{\Phi}\rangle=1/4$ (i.e., by artificially removing cavity backaction). The curves in (b) are found for the corresponding parameters by diagonalizing the full quantum model  of Eq. (\ref{Hamil_BH:1D}). The other parameters are as described in Fig.~\ref{Fig:N_mu_Dc}. } \label{Fig:1D_tzero}
\end{figure}

It is interesting to compare the results for our system with the predictions for a one-dimensional bichromatic lattice with incommensurate wave lengths \cite{Roux08,Deng08}. For this purpose we substitute the operator $\hat{\Phi}$ in Eq.\ \eqref{Hamil_BH:1D} by a scalar $\hat{\Phi}\to 1/4$. This choice is made in order to obtain similar curves at commensurate densities $\bar n=0,1,2$ for $s_0=0.004\kappa$. Figure~\ref{Fig:1D_tzero}(a) displays the resulting density as a function of the chemical potential for different strengths of the cavity field and $\delta_c<0$. We observe that, by increasing $s_0$ in the model where  $\hat\Phi$ is a scalar, the parameter regions for which the particle density is constant and hence the gas is incompressible, rapidly shrink \cite{Footnote:plateaus}. This trend is significantly slower for the case in which cavity backaction is taken into account, as can be observed in Fig.~\ref{Fig:1D_tzero}(b). In addition, in presence of cavity backaction, discontinuities in the values of the compressibility are encountered and seem to correspond to the first-order phase transition \cite{Footnote:QMC:2}.  This behaviour qualitatively differs from the one encountered when artificially removing the effect of the cavity, as shown in Fig. \ref{Fig:1D_tzero}(a). For the largest value of the laser intensity considered here, $s_0=0.008\kappa$, the incompressible phases disappear.

\subsubsection{Phase diagram for $t>0$}

%{\bf Figures from Andr\'e:}

In the following we discuss our results for $t>0$, which we obtained with a quantum Monte Carlo (QMC) approach \cite{Batrouni:1,Batrouni:2}, whose details are reported in the Appendix~\ref{AppQMC}. Figure~\ref{fig:coeff008}(a)  displays the averaged density versus chemical potential for $s_0/\kappa = 0.008$ and increasing values of the tunneling rate. For $t\to 0$ one recovers the black curve in Fig.~\ref{Fig:1D_tzero}(b), while as $t$ grows the curve is increasingly shifted towards negative values of $\mu$, and tends to a continuous line at larger values of $\mu$. The discontinuities in the mean density and in the cavity field amplitude observable in Fig. ~\ref{fig:coeff008} lead to a step-like dependency of the compressibility from $\mu$. The three vertical bars for $t=0.096 U$ indicate the three values of $\mu/U$ which we choose for plotting the density profile, $\langle n_i\rangle$ vs. $i$, in Fig.~\ref{fig:density008}. Here, we observe that as $\mu$ is increased the amplitude of the density oscillations increases. The appearance of clusters where the density has periodicity $\lambda_0=2d_0$ is due to the fact that $\lambda\sim 2d_0$: These oscillations locally maximize the value of $\langle\hat\Phi\rangle$. Since the ratio $\lambda/d_0$ is incommensurate (for the system size we consider), this pattern can only exist locally: The size of the clusters is in fact limited by the beating between the two spatial periodicities. In particular, after a number of sites of the order of the length scale of the beating signal, the density fluctuations increase and allow the atomic distribution to reorganize in the external potential. In this way, the fields emitted by the atoms add up coherently and the intracavity field is maximized. In all considered cases the onsite energies exceeds the onsite repulsion.

\begin{figure}
\includegraphics[width=8cm]{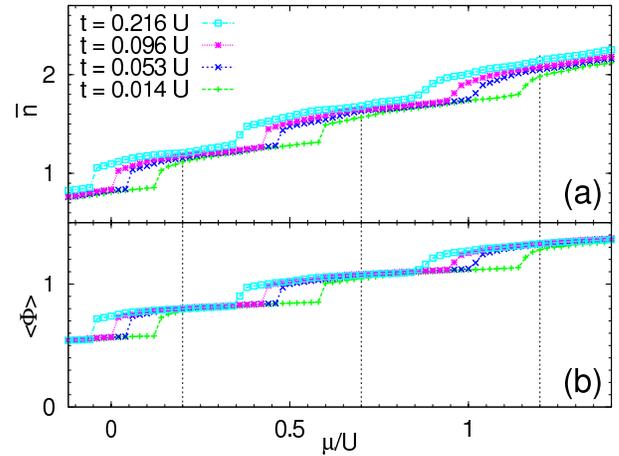}
\caption{(color online) (a) Mean density $\bar{n}$ and (b) expectation value $\langle\hat\Phi\rangle$ versus the chemical potential $\mu$ (in units of $U$) for a 1D lattice at different tunneling values $t/U=0.014,0.053,0.096,0.216$. The other parameters are the same as for the black line at $s_0=0.008\kappa$ in Fig.~\ref{Fig:1D_tzero}(b). The curves have been evaluated by means of a QMC program described in Appendix \ref{AppQMC}. The vertical dotted lines indicate the values of $\mu$ in Fig. \ref{fig:density008}. }\label{fig:coeff008}
\end{figure}

\begin{figure*}
\includegraphics[width=0.33\textwidth]{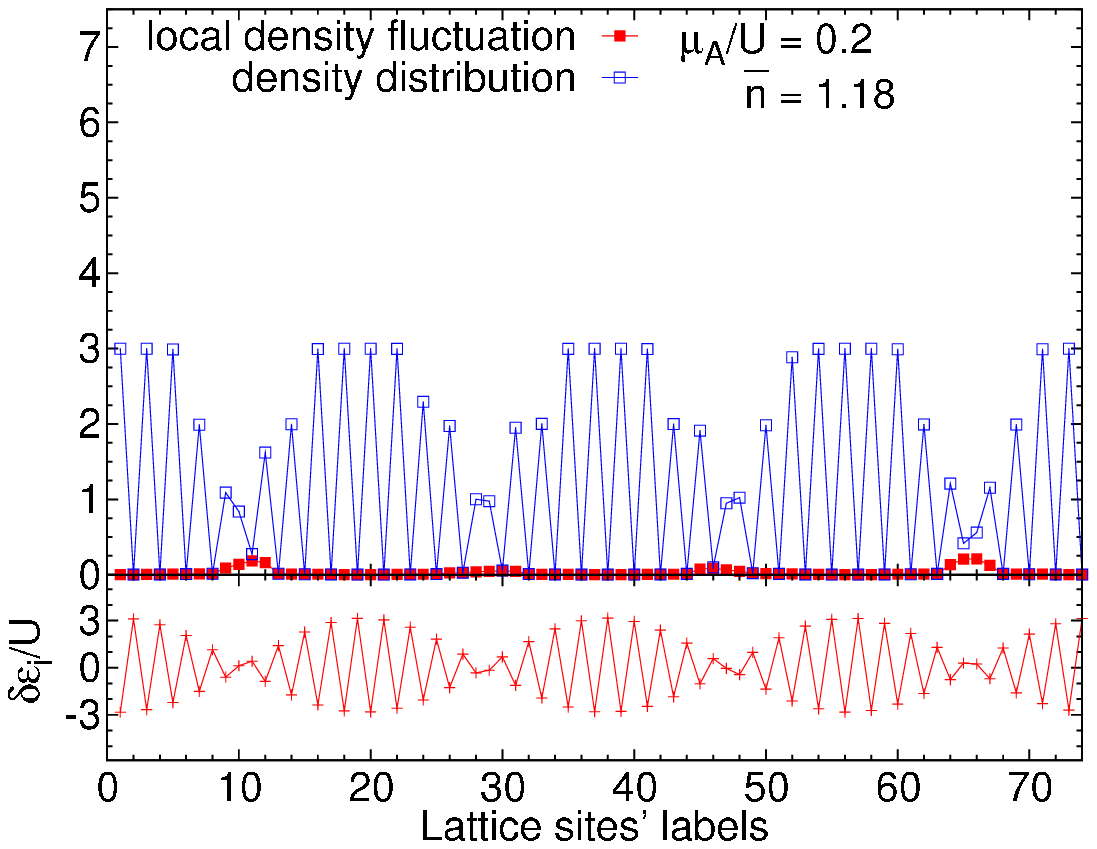}
\includegraphics[width=0.33\textwidth]{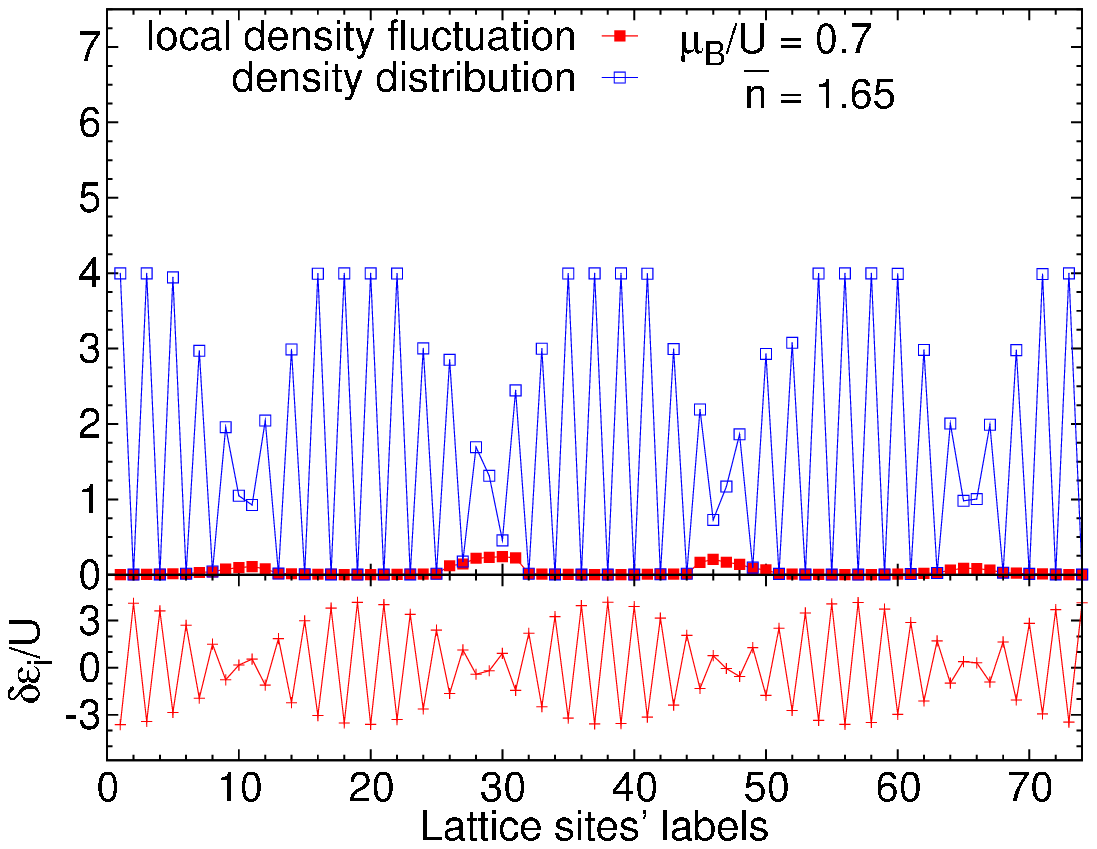}
\includegraphics[width=0.33\textwidth]{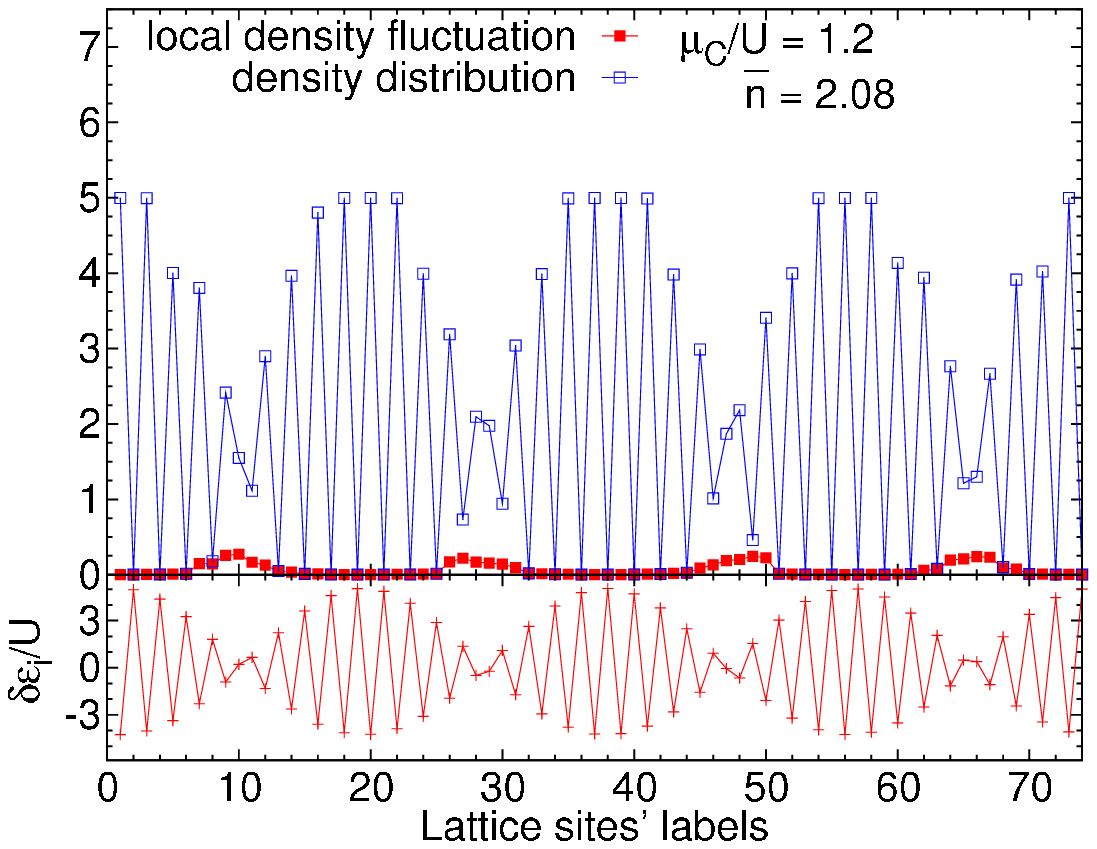}
\caption{(color online) Upper row: Onsite density distribution $\langle\hat{n}_i\rangle$ and local density fluctuations $\langle\hat{n}_i^2\rangle-\langle\hat{n}_i\rangle^2$ as a function of the site $i$. The line connecting the points serves as guide for the eyes. Lower row: onsite energy due to the cavity field, $\delta \epsilon_i$ (in units of $U$). The curves are evaluated for the parameters of the curve in Fig.~\ref{fig:coeff008} with $t=0.096 U$ and $s_0=0.008\kappa$. The plots correspond to $\mu/U=0.2, 0.7, 1.2$ (from left to right).}\label{fig:density008}
\end{figure*}

\begin{figure} 
\includegraphics[width=.45\textwidth]{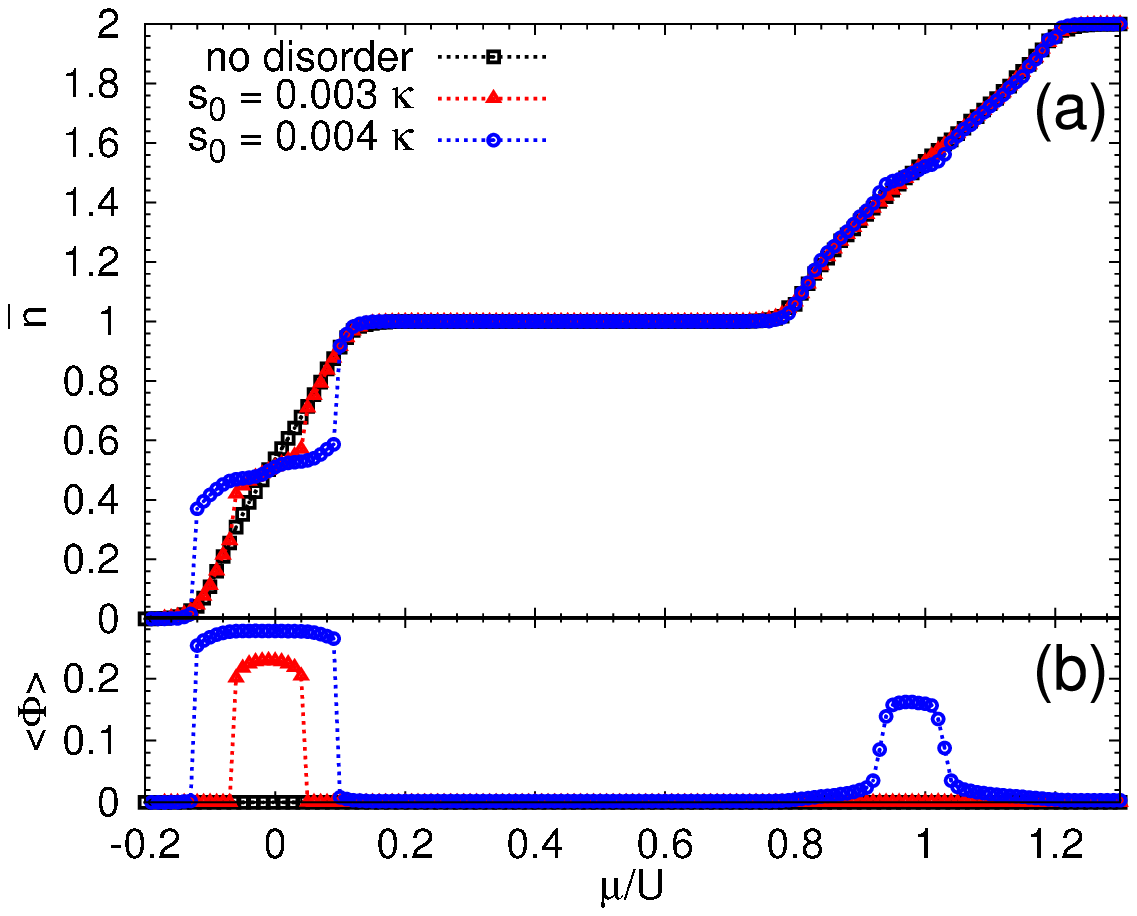}
\includegraphics[width=.45\textwidth]{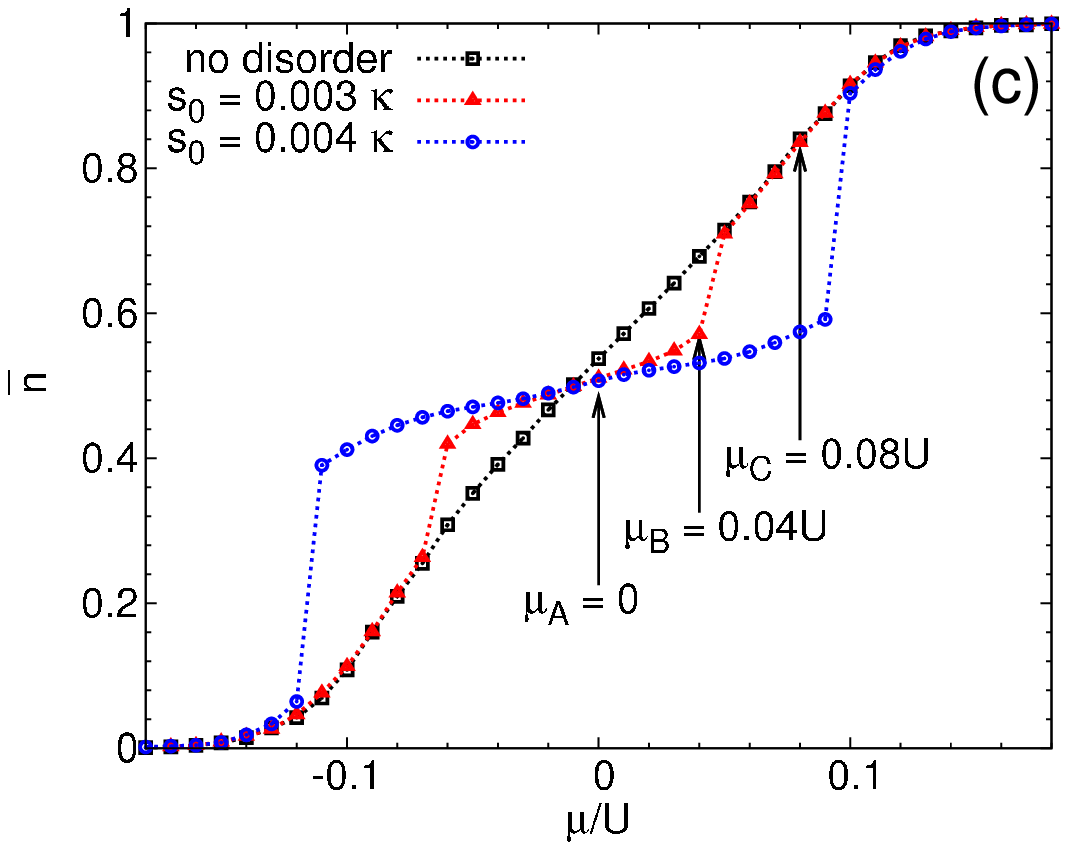}
\includegraphics[width=.45\textwidth]{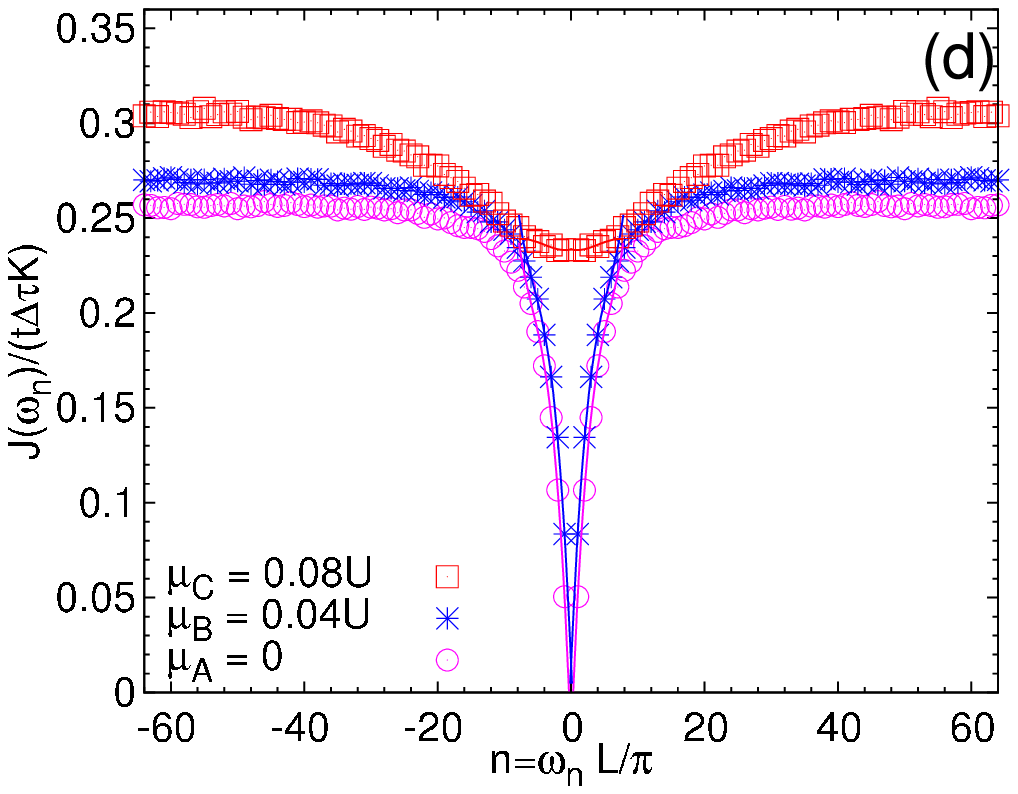}
\caption{(color online) Results of QMC simulations for a 1D lattice with 74 sites and $t=0.053$ with periodic
boundary conditions. (a) Mean density \(\bar{n}\) and (b) modulus of the mean intracavity field amplitude $\langle\hat{\Phi}\rangle$ versus $\mu$ (in units of $U$) for \(s_0 = 0.003 \kappa\) (triangles), \(s_0 = 0.004 \kappa\) (circles) and in the absence of the pump laser \(s_0=0\) (squares). (c) Zoom of mean density in the region of parameters with $-0.18\le \mu\le 0.18$. (d) The Fourier transform of the pseudo current-current correlation function $\tilde{J}(\omega)$ for the values of $\mu/U$ indicated by the arrows in (c) for $s_0=0.003\kappa$. The used simulation parameters are $L=128$ and $\Delta \tau=1$.The lines shows cubic interpolations of the data. The extrapolated value at zero frequency is the estimation of the SF density. The other parameters are the same as in Fig.~\ref{fig:coeff008}.\label{fig:rho_mu}
}
\end{figure}

Figures~\ref{fig:rho_mu}(a) and (b) display $\bar{n}$ and $\langle\hat\Phi\rangle$, which is proportional to the cavity field amplitude, as a function of $\mu$. The curves are evaluated at $t=0.053 U$ and for two different values of $s_0$, such that the onsite energy is typically smaller than the onsite repulsion. The case corresponding to the typical Bose-Hubbard model, which is found here by setting $s_0=0$ (i.e., the pump laser is off), is reported for comparison. In presence of the laser incompressible phases are still found. For the corresponding values of the chemical potential the cavity field vanishes. We observe, in addition, jumps in the density between commensurate values of $\bar{n}$, for which the intracavity photon number is different from zero. Figure~\ref{fig:rho_mu}(d) reports the Fourier transform of the pseudo current-current correlation function $J(\omega)$ \cite{Batrouni:1,Batrouni:2}  for three points of the phase diagram, which are indicated by the arrows in Fig.~\ref{fig:rho_mu}(c) (details of the procedure are reported in Appendix~\ref{AppQMC}). The superfluid (SF) density is obtained by the extrapolated value at zero frequency, $\omega=0$. In the whole parameter region where the cavity field is different from zero, we observe that the SF density vanishes. This behavior, together with the non-vanishing compressibility, is characteristic of a Bose-glass phase. Outside this region the gas is SF.

\begin{figure*}
\includegraphics[width=0.33\textwidth]{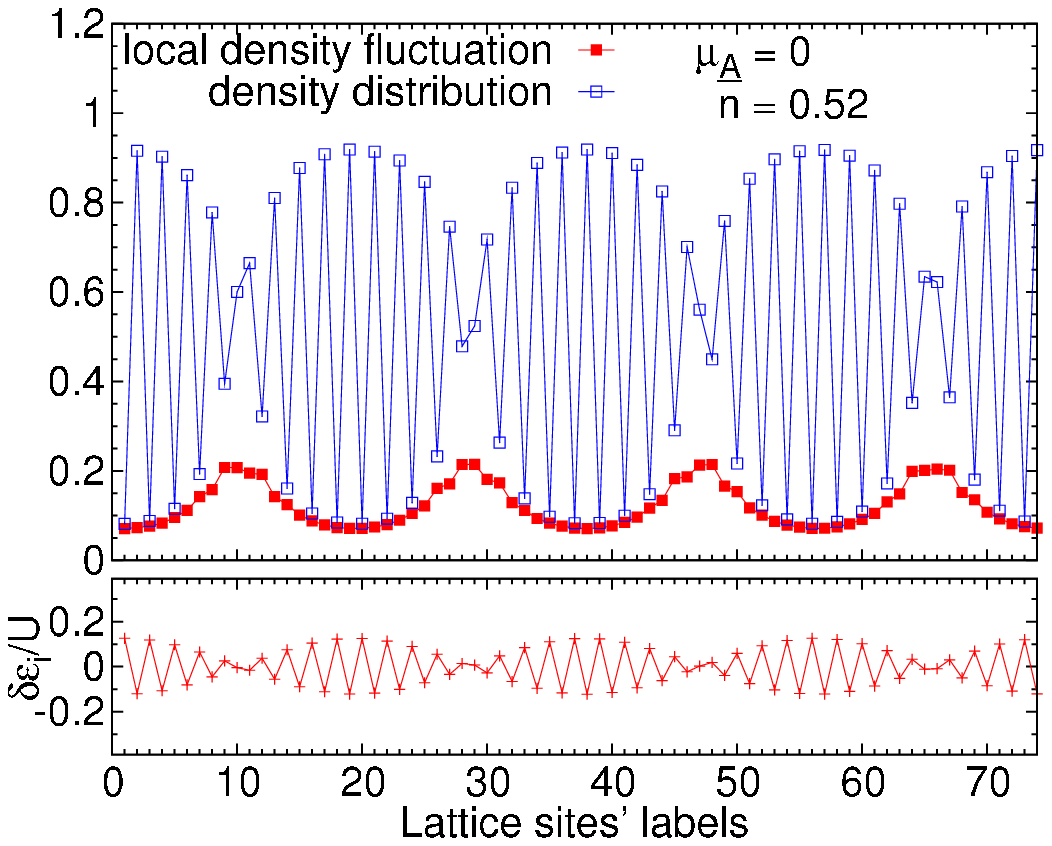}
\includegraphics[width=0.33\textwidth]{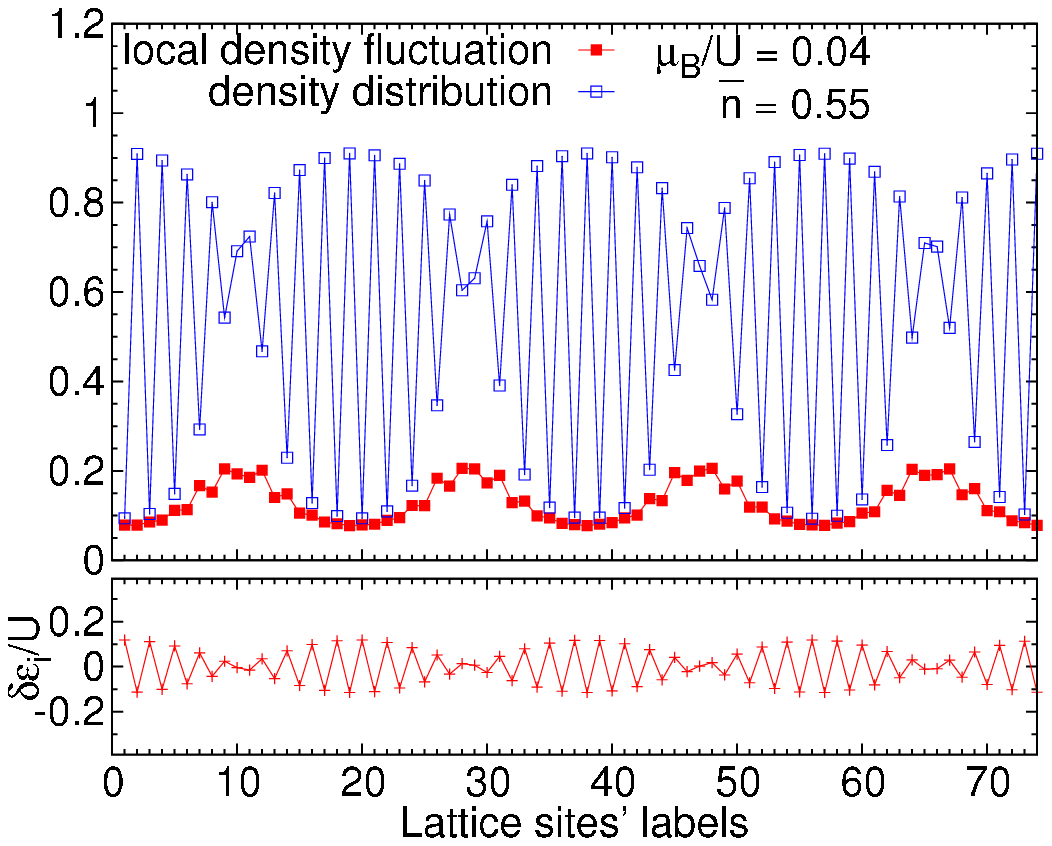}
\includegraphics[width=0.33\textwidth]{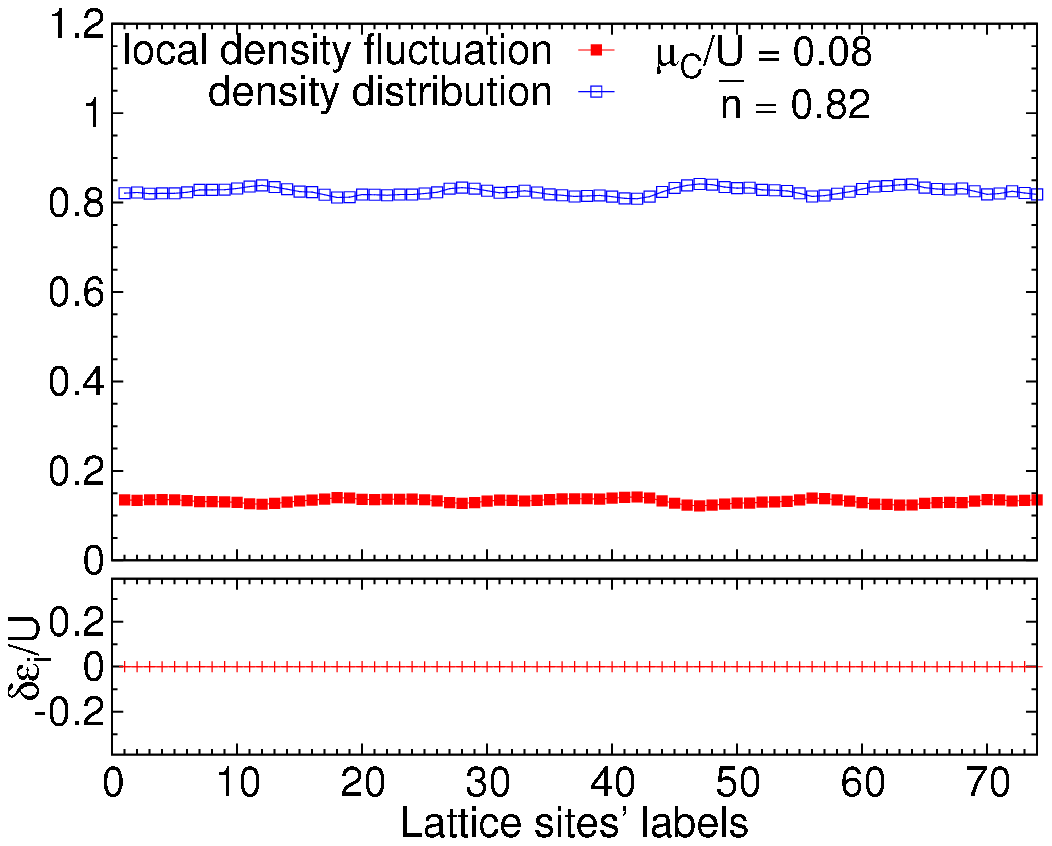}
\caption{(color online)  Upper row: Onsite density distribution $\langle\hat{n}_i\rangle$ and the local density fluctuation $\langle\hat{n}_i^2\rangle-\langle\hat{n}_i\rangle^2$ as a function of the site $i$. The line connecting the points serves as guide for the eyes. Lower row: onsite energy due to the cavity field, $\delta \epsilon_i$ (in units of $U$). The curves are evaluated for the parameters of the red curve in Fig.~\ref{fig:rho_mu} with $t=0.053 U$ and $s_{0}=0.003\kappa$. The plots correspond to $\mu/U=0, 0.04, 0.08$ (from left to right).}\label{fig:density003} 
\end{figure*}

The density profile, $\langle n_i\rangle$ versus $i$, is shown in Fig.~\ref{fig:density003} for the three values of $\mu$ indicated in Fig. \ref{fig:rho_mu}(c). In the left plot, corresponding to vanishing SF fraction, we observe a density modulation and thus localized density fluctuations. These become less localized for $\mu=0.04$, while in the SF region, where the number of photons is zero, the density is uniform along the lattice. 

\begin{figure}
\includegraphics[width=.45\textwidth]{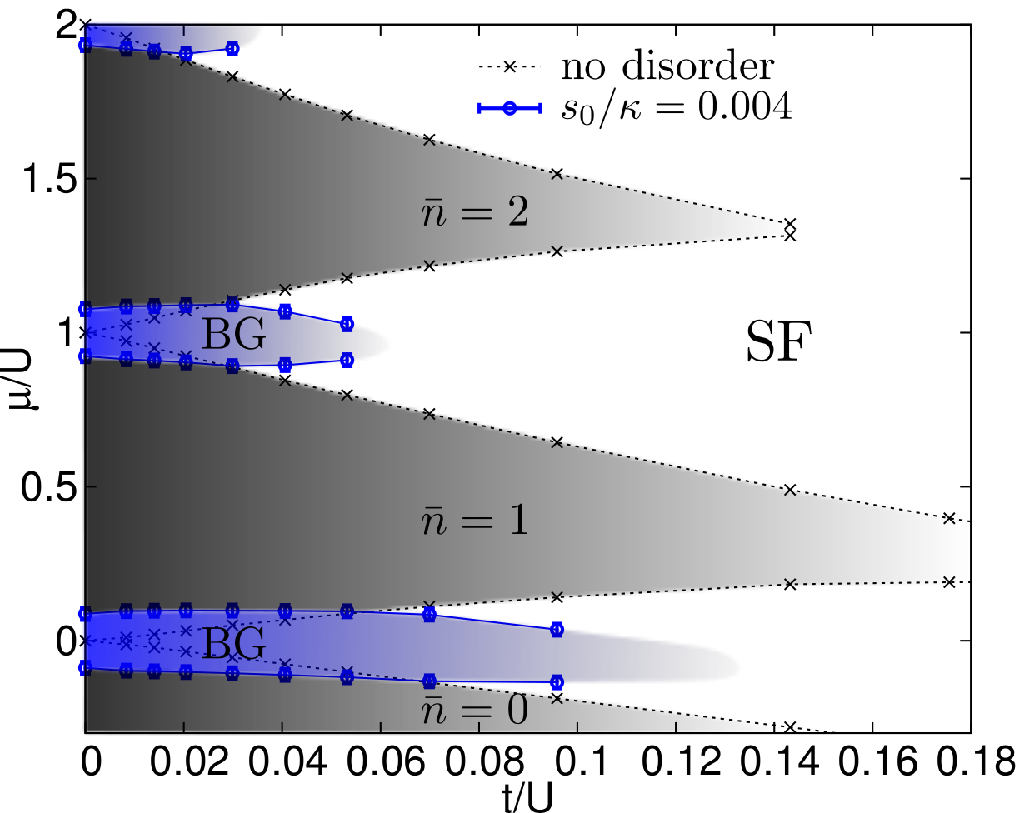}
\caption{(color online) The phase diagram is obtained by QMC calculation for a 1D lattice with 74 sites for $s_0=0.004\kappa$. The other parameters are as in Fig.~\ref{fig:rho_mu}. The results are compared with the pure case (dotted curves).}\label{fig:phase_diagram_1D}
\end{figure}

The phase diagram is extrapolated by tracking the behavior of the density $\bar{n}$ versus the chemical potential for different tunneling values. Figure~\ref{fig:phase_diagram_1D} displays the resulting phase diagram in the $\mu-t$ parameter plane. The grey regions indicate the MI states at densities $\bar n=0,1,2$, the blue regions a compressible phase with vanishing SF density, where the number of intracavity photon is large, while outside these shaded region the phase is SF. The effect of cavity backaction is evident at low tunneling, where $\langle\hat\Phi\rangle>0$: Here the size of the MI regions is reduced and one observes a direct transition between MI and Bose-glass (BG) phase. At larger tunneling a direct MI-SF transition occurs and the MI-SF phase boundary merges with the one found for $s_0=0$: In fact, for larger quantum fluctuations $\langle\hat\Phi\rangle\to 0$ in the thermodynamic limit. This feature is strikingly different from the situation in which the incommensurate potential is classical \cite{Roux08,Deng08}, where, the MI lobes shrink at all values of $t$ with respect to the pure case. 
Before we conclude this section, some remarks regarding the determination of the phase diagram must be made. In fact, we have used the grand-canonical ensemble to obtain the phase diagram in presence of cavity backaction. In the absence of the cavity backaction ($\Phi\to0$) we have used the canonical approach for QMC simulation, which allows us an accurate determination of the transition borders. 

\subsection{Two-dimensional lattice}
\label{sec:results:MF}

We now analyze the phase diagram of a two-dimensional (2D) lattice, of which one axis coincides with the cavity axis while in the perpendicular direction the atoms are pumped by the standing wave laser which is quasi-resonant with the cavity field. In this situation, hence, the site-dependent term proportional to the laser intensity ($V_1$) in Eq. \eqref{DeltaMu} is relevant. The presence of this classical field, in fact, significantly affects the phase diagram even in the absence of the cavity, since its wavelength is taken to be incommensurate with the periodicity of the confining optical lattice.

Before we discuss the results, some consideration on the parameters is in order. For the parameters we take the effect of the classical incommensurate potential proportional to $V_1$ dominates over the cavity incommensurate field in determining the value of the onsite energy, Eq. \eqref{DeltaMu}. In particular, the sign of the coefficient $V_1$ is determined by the detuning $\Delta_a$. The size of the lattice is fixed to vary about the value $K\sim 300\times 300$, then the expectation value of operator $\hat\delta_{\rm eff}$ in Eq. \eqref{delta:eff} is such that it can be approximated by $\hat\delta_{\rm eff}\sim -u_0 \sum_{i,j}Y_{0}^{(i,j)}\hat{n}_{i,j}/K$. This property shows that the sign of the cavity-induced potential in the Bose-Hubbard Hamiltonian, Eq. \eqref{Hamil_BH:1}, is controlled by the sign of $u_0$, and thus of the detuning $\Delta_a$ between atom and pump. A simple check of the sign of the coefficients in Eq. \eqref{Hamil_BH:1} shows that the formation of finite intracavity potentials is energetically favoured when $\Delta_a>0$. 

\begin{figure}
\centering
\includegraphics[width=.45\textwidth]{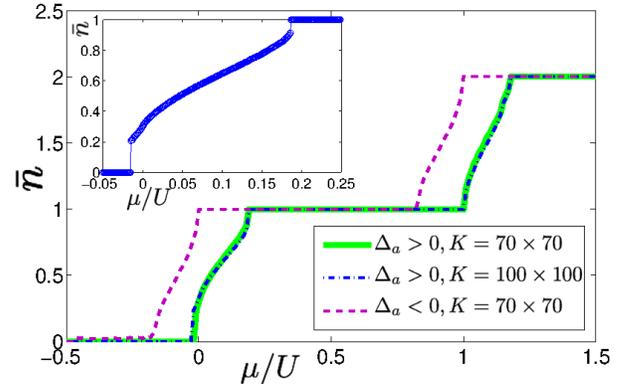}
\caption[]{(color online) Mean density $\bar{n}$ versus $\mu$ (in units of $U$) for a 2D lattice. The curve is evaluated by exact diagonalization of Hamiltonian \eqref{Hamil_BH} for $t=0$ and $K=70\times 70$ and $\Delta_a<0$ (dashed line),  $K=70\times 70$ and $\Delta_a>0$ (solid line). The parameters are $|\Delta_a|=2\pi\times 58 GHz$,  $s_0=0.15\kappa$, $\delta_c=-5\kappa$, $u_0=237\kappa$ ($\kappa=2\pi \times 1.3$ MHz). Inset: Zoom of the curve at $K=70\times 70$ and $\Delta_a>0$ in the compressible phase. The dashed-dotted line for $K=100\times 100$ and $\Delta_a>0$ shows, when compared with the solid line, that the qualitative behaviour of the curves remain invariant as the system size is scaled up.} \label{Fig:2D_tzero_u0}
\end{figure}

We first analyse the behaviour of the mean density as a function of the chemical potential for $t\to 0$ for opposite signs of $\Delta_a$, which is found by determining the ground state of the two-dimensional Hamiltonian in Eq \eqref{Hamil_BH:1} after setting the tunneling coefficient $t=0$. The results are displayed in Fig. \ref{Fig:2D_tzero_u0}. For the considered set of parameters we observe the appearance of incompressible phases. To a very good approximation they are in the interval of values determined by the classical incommensurate potential $V_1$, which takes either positive or negative values depending on whether $\Delta_a$ is positive or negative. For commensurate density $\bar n=1$ and $\Delta_a<0$,  for instance, an incompressible phase is found in the interval $0\lesssim\mu\leq\mu_1<U$, where $\mu_1$ depends on $V_1$. For  $\Delta_a>0$, instead, the incompressible phase is in the interval $0<\mu_1\leq\mu\lesssim1$. Different from the 1D case, the incommensurate phase shrinks only on one side due to the effect of the classical pump (while the cavity potential is a small correction). The dominant effect of the classical field is also visible when analysing the curve in the parameter regime where the phase is compressible: The inset shows a zoom of the curve for $\Delta_a>0$, which exhibits various discontinuities in the compressibility. The finite compressibility is mostly due to the classical field. The jump of the density at $\mu'$ (with $-0.05<\mu'<0$) corresponds to the region in which an intracavity potential is built and is thus due to cavity backaction. 
Note that this discontinuity is observed at a shifted value of $\mu$ when $\Delta_a<0$.

We now determine the behaviour at finite $t$ for the 2D lattice taking $\Delta_a>0$ by means of a local mean-field calculation. This is performed by setting $\hat b_{i,j}=\psi_{i,j}+\delta \hat{b}_{i,j}$ where $\psi_{i,j}=\langle \hat b_{i,j} \rangle$ is a scalar giving the local SF order parameter and $\delta \hat{b}_{i,j}$ are the fluctuations with zero mean value. The new form is substituted in Eq. \eqref{Hamil_BH} and the second order fluctuations of the hopping term, namely, the terms  $\delta\hat{b}_{i,j}\,\delta\hat{b}_{i',j'}$, are discarded \cite{Sheshadri}. The resulting Bose-Hubbard Hamiltonian in the mean-field approximation 
takes the form $\hat{\mathcal H}_{\rm BH}^{(MF)}=\sum_{i,j}\hat{\mathcal H}_{i,j}$, 
where 
\begin{eqnarray}\label{H_MF}
\hat{\mathcal H}_{i,j}=-t \eta_{i,j}\left(\hat b_{i,j}^\dag-\frac{\psi_{i,j}^\ast}{2}\right)+{\rm H.c.}+ \frac{U}{2} \hat n_{i,j}(\hat n_{i,j}-1)+\hat\epsilon_{i,j} \hat n_{i,j}\,,
\end{eqnarray}
and $\eta_{i,j}=\psi_{i+1,j}+\psi_{i-1,j}+\psi_{i,j+1}+\psi_{i,j-1}$ is the sum of the local SF parameters of the neighbouring sites. We remark that cavity back-action makes Hamiltonian $\hat{\mathcal H}_{i,j}$ in (\ref{H_MF}) non-local in the density, since it depends on the collective operator $\hat\Phi$ appearing in $\hat\epsilon_{i,j}$. The local SF order parameters $\psi_{i,j}$ are found by solving the coupled set of self-consistency equations $\psi_{i,j}=\langle \phi_G^{(MF)}|\hat{b}_{i,j}|\phi_G^{(MF)}\rangle$, where $|\phi_G^{(MF)}\rangle=\otimes_{i,j=1}^K|\phi_{i,j}\rangle$ is the ground state in the mean-field approximation. It is thus the direct product of the single-site states $|\phi_{i,j}\rangle$, defined as
\begin{equation}
|\phi_{i,j}\rangle=\sum_{n=0}^\infty \alpha_n^{(i,j)} |n\rangle_{i,j}\,,
\end{equation}
%{\bf GM: Hessam give please the definition you took.}
in which $|n\rangle_{i,j}$ is the state of $n$ bosons at a lattice site with $(x_i,z_j)$, and $\sum_{i,j}|\alpha_n^{(i,j)}|^2=1$.
In our numerical implementation the evaluation of ground state is repeated till the averaged SF order parameter $\psi=\sum_{i,j}\psi_{i,j}/K$ converges up to a tolerance of 0.005. The recursive calculation of the ground states  of the self-consistent Hamiltonian $\hat{\mathcal H}_{\rm BH}^{(MF)}$ is terminated once the value of $\bar{n}$ converges with an accuracy of $2\times10^{-4}$.

\begin{figure}
\centering
\includegraphics[width=.47\textwidth]{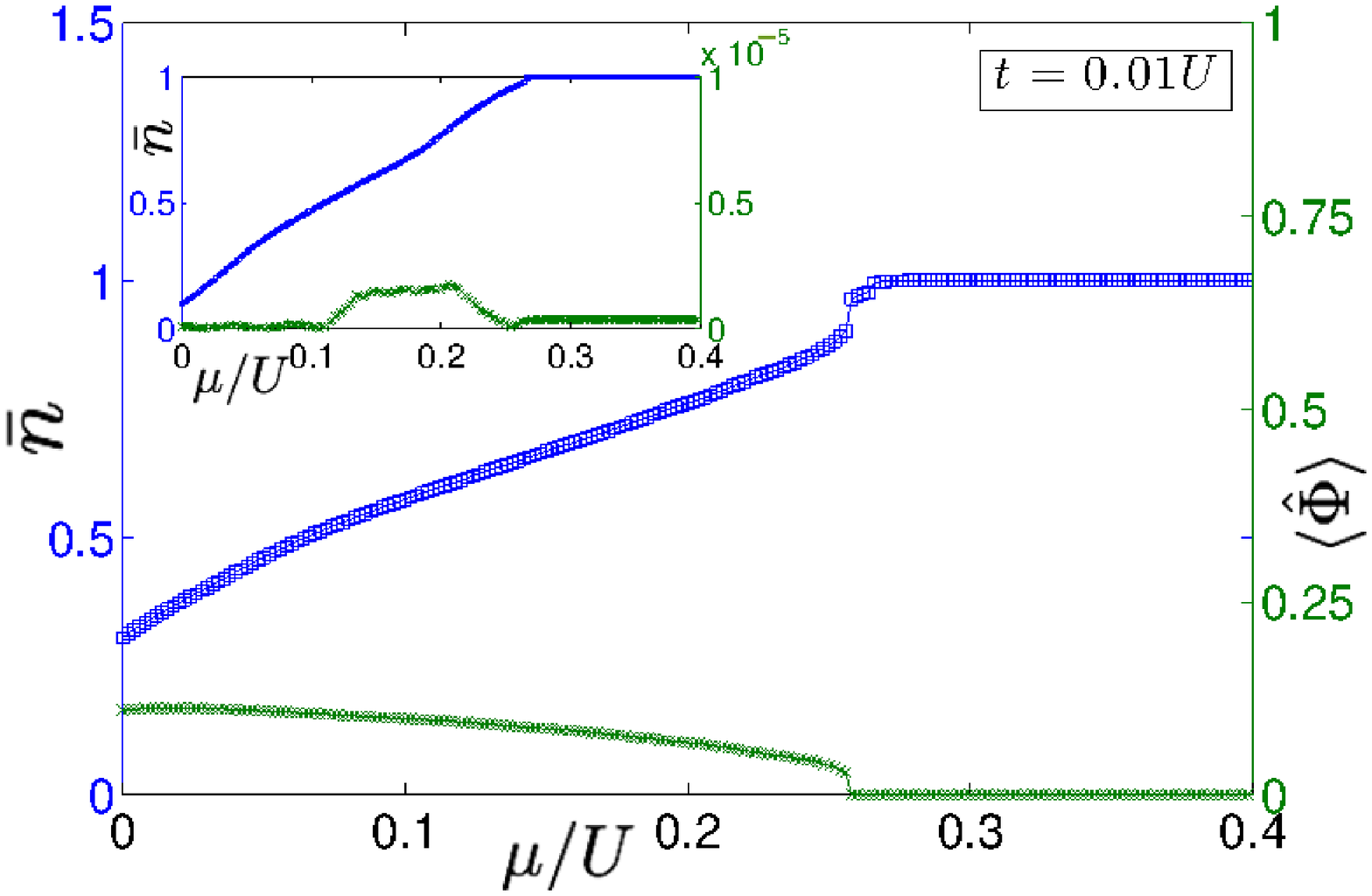}
\caption[]{(color online) Mean density $\bar{n}$ (blue line) and $\langle \hat\Phi\rangle$ (green line with crosses) versus $\mu$ (in units of $U$) for a 2D lattice. The curves are evaluated using local mean field for $t=0.01U$ and $K=70\times 70$.  The parameters are $\Delta_a=2\pi\times 58$~ GHz,  $s_0=0.15\kappa$, $\delta_c=-5\kappa$, $u_0=237\kappa$ ($\kappa=2\pi \times 1.3$ MHz). Inset: Same curves but for $\delta_c=-300\kappa$. Note that the maximum value of $\langle \hat\Phi\rangle$ is almost 5 orders of magnitude smaller than for $\delta_c=-5\kappa$.} \label{Fig:MF_n_mu}
\end{figure}

Figure~\ref{Fig:MF_n_mu} displays the mean density as a function of the chemical potential for the same parameters of the solid curve in Fig. \ref{Fig:2D_tzero_u0} but for $t=0.01U$. The zoom on the region of parameters where the compressibility is different from zero shows that also at finite $t$ the curve is discontinuous. The jumps indicate the interval of values in which there is an intracavity field (see crosses).  The inset displays the corresponding curve when the pump is far detuned from the cavity field: the compressibility does not present jumps in the compressible phase and the mean intracavity field is at least three orders of magnitude smaller.

\begin{figure}
\centering
\includegraphics[width=.44\textwidth]{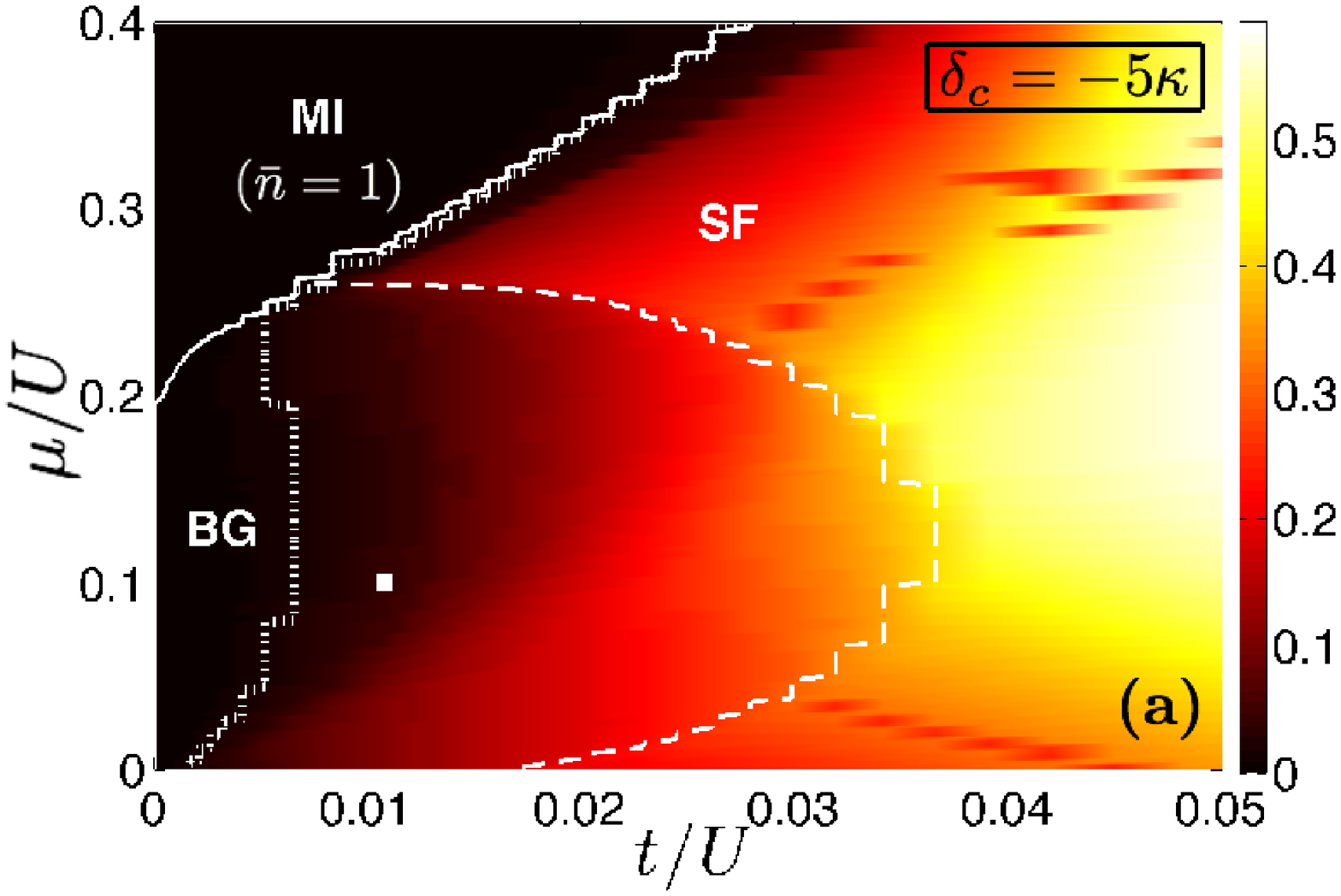}
\includegraphics[width=.44\textwidth]{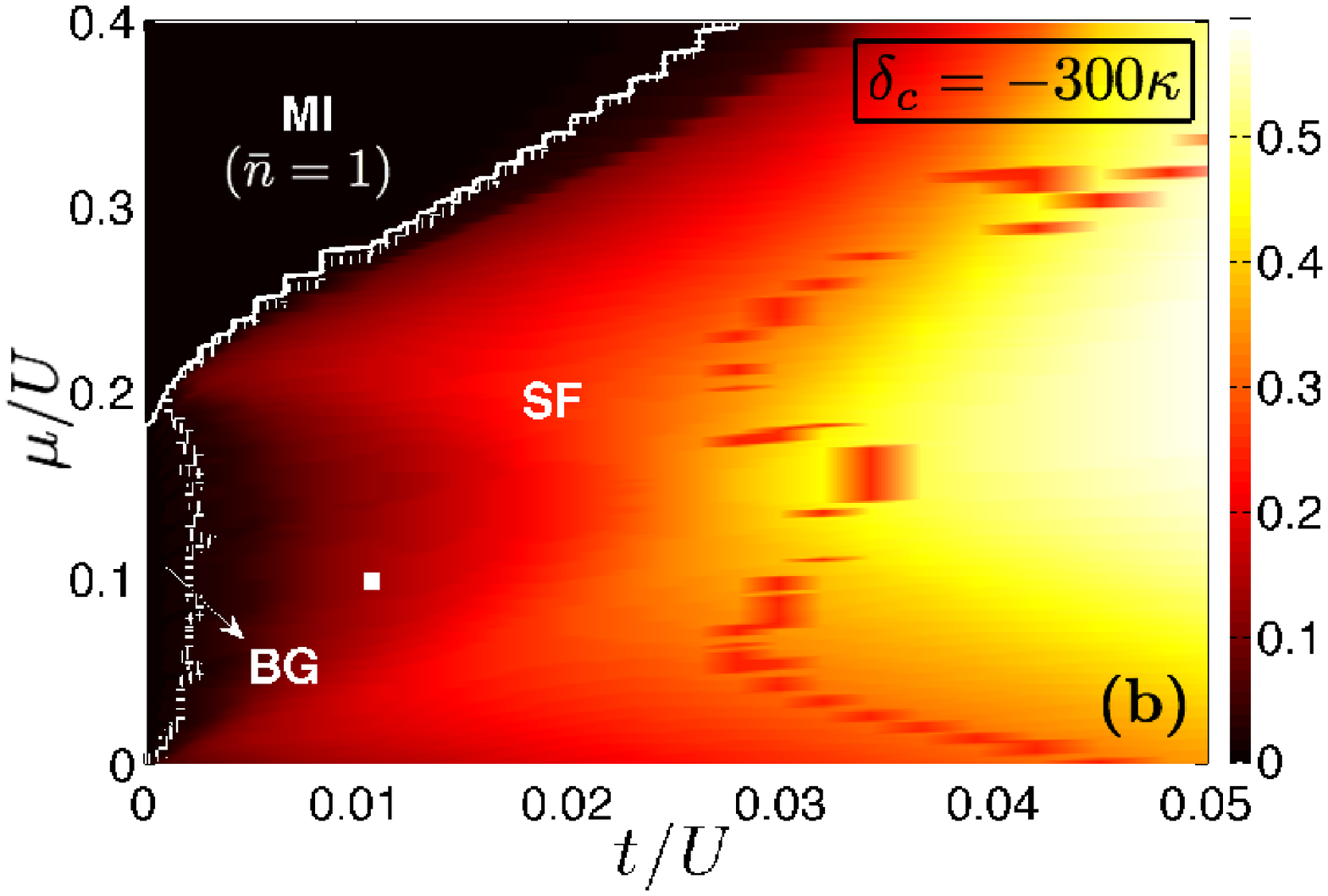}
\includegraphics[width=9.2cm]{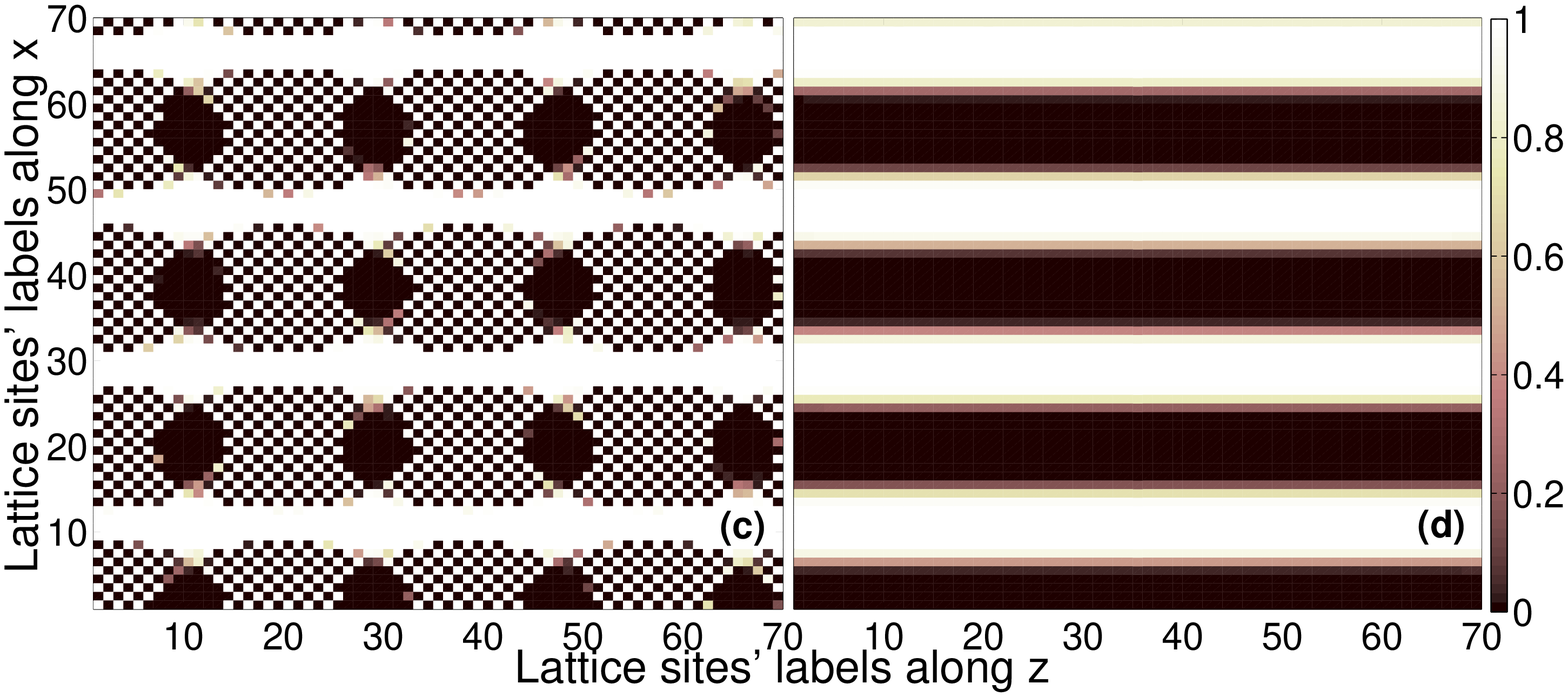}
\caption[]{(color online)
(a), (b) Order parameters in $\mu$-$t$ plane (in units of $U$) obtained by the mean-field calculation for a $70\time70$ lattice with periodic boundary conditions. The dotted lines separate the region with vanishing order parameters, while the solid line identifies the border for the incompressible MI state at density $\bar{n}$. The regions with finite compressibility and vanishing order parameters correspond to BG phases. The dashed line separates the region where the photon number is 2 order of magnitude larger than outside. The parameters are $s_0=0.15\kappa$, $u_0=237\kappa$, $\Delta_a=2\pi\times 58$ GHz, whereas (a) $\delta_c=-5\kappa$ and (b) $\delta_c=-300\kappa$. In the latter case the effect of the cavity potential is expected to be small. The local densities $\langle\hat{n}_{i,j}\rangle$ of the phase diagram at $\mu=0.1 U$ and $t=0.01\kappa$ are shown in (c) for $\delta_c=-5\kappa$ and $\bar{n}=0.57$ (squared point in (a)) and in (d) for $\delta_c=-300\kappa$ and $\bar{n}=0.47$ (squared point in (b)).} \label{Fig:PhaseDiag1}
\end{figure}

Figure~\ref{Fig:PhaseDiag1}(a) displays the mean SF order parameter in the $\mu-t$ plane and for density $\bar{n}\le 1$. Here, the dotted lines identify the regions where the order parameter takes values below 0.02. The solid curve indicates where the gap in the spectrum is different from zero, corresponding to vanishing density fluctuations $\Delta\varrho= (\overline{n^2}-\overline{n}^2)^{1/2}$, where  $\overline{n}=\sum_{i,j}\langle  \hat{n}_{i,j}\rangle/K$ and $\overline{n^2}=\sum_{i,j}\langle  \hat{n}^2_{i,j}\rangle/K$ (the threshold is set at 0.02). For comparison, Fig.~\ref{Fig:PhaseDiag1}(b) displays the corresponding diagram when the cavity is pumped far off-resonance, so that the effect of cavity backaction is very small and practically negligible. We note that the curve delimiting  the MI phase has a very similar behaviour in presence and in absence of cavity back-action, showing that for the considered parameters the existence of incompressible phases is determined by the transverse optical lattice. The behaviour of the compressible phase with vanishing order parameter, which we here denote by BG phase, varies instead significantly in presence of the cavity potential, as one can observe by comparing Fig.~\ref{Fig:PhaseDiag1}(a) and (b). We finally point out the region delimited by the dashed line, which appears only in the subplot (a): This indicates the parameters for which the mean value of $\hat\Phi$ is at least two orders of magnitude larger than outside. In this region there is an intracavity field, which is due to coherent scattering by the atoms. 

The typical onsite density encountered in this parameter region, and in particular for the parameters indicated by the squared point in Fig.~\ref{Fig:PhaseDiag1}(a), is shown in subplot (c). We compare it with the case without cavity backaction: the density corresponding to the squared point in Fig.~\ref{Fig:PhaseDiag1}(b) is displayed in subplot (d). Without cavity backaction one observes dark stripes along the vertical direction at which the density is minimum. The stripes are almost regularly distributed and are due to the classical incommensurate potential along the $x$ axis.  When cavity backaction becomes relevant, an incommensurate potential also appears along the $z$ direction. This intracavity potential is associated with the appearance of clusters within which the density exhibit a checkerboard distribution, as shown in Fig.~\ref{Fig:PhaseDiag1}(c). These clusters are the two-dimensional analogy of the density-wave like behaviour observed in 1D: they maximize scattering into the cavity field and their size is determined by the length due to the beating between the lattice wave length and the incommensurate cavity potential.

\begin{figure}
\centering
\includegraphics[width=.45\textwidth]{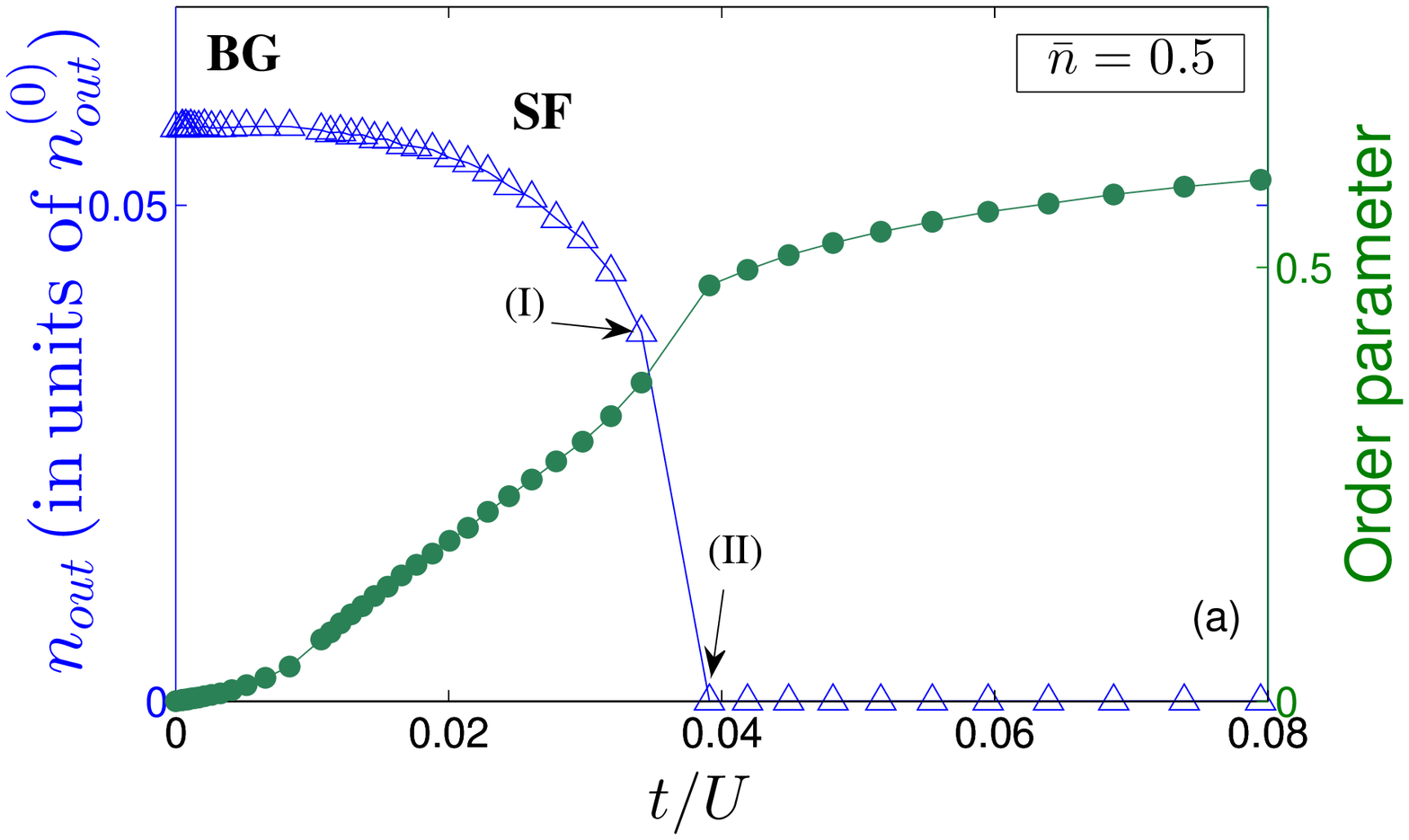}
\includegraphics[width=9cm]{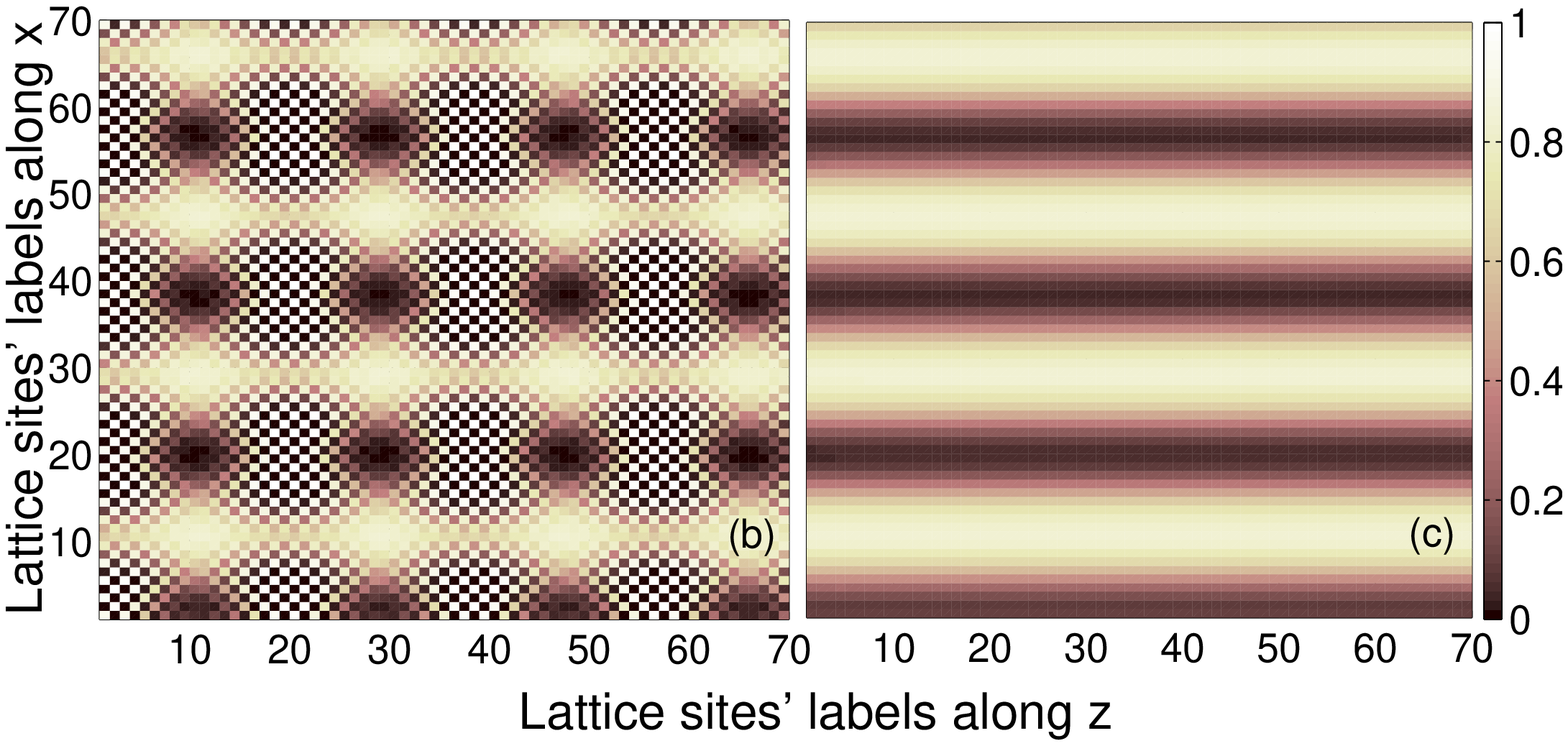}
\includegraphics[width=.43\textwidth]{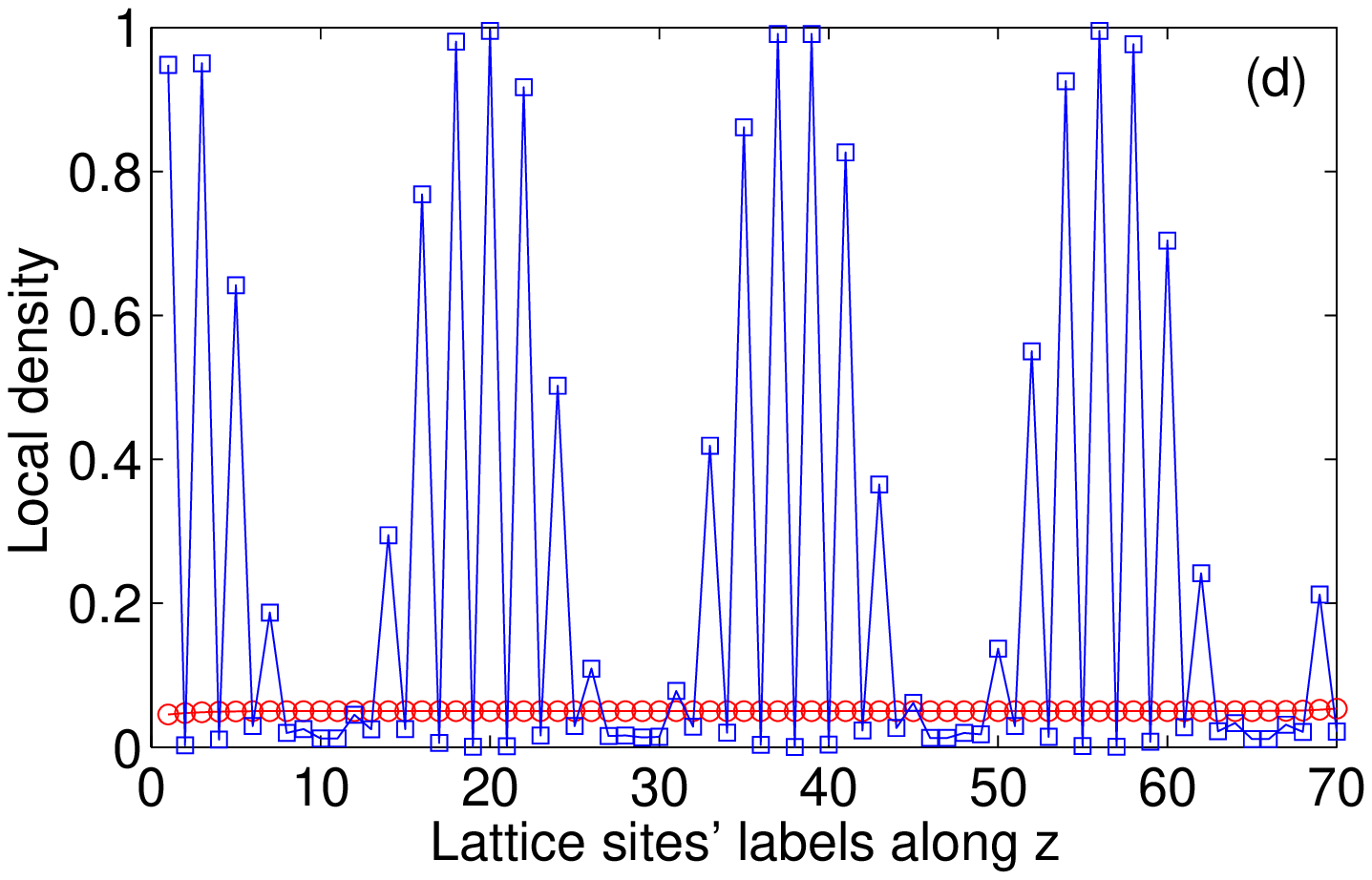}
\caption[]{(color online) (a) Intensity at the cavity output, $n_{\rm out}=\langle a_{\rm out}^\dag a_{\rm out}\rangle$ 
as a function of $t$ (in units of $U$) for the parameters of Fig.~\ref{Fig:PhaseDiag1}(a) and by fixing $\bar{n}=0.5$.
Here, $n_{\rm out}$ is in units of $n_{\rm out}^{(0)}=\kappa n_{\rm cav}^ {\rm (max)}$ where $n_{\rm cav}^ {\rm (max)}=s_0^2K/(\hat\delta_{\rm eff}^2+\kappa^2)$ is the maximum number of intracavity photons, obtained when all atoms scatter in phase into the cavity mode. The curve with circles (right $y$-axis) gives the corresponding order parameter. Subplot (b) displays the countour plot of the  local density distributions at point (I) in panel (a), where $t=0.034 U$, $\mu=0.106 U$, $\langle\hat\Phi\rangle=0.136$. Subplot (c) displays the local density distributions at point (II) in panel (a), where $t=0.039 U$, $\mu=0.122 U$, while $\langle\hat\Phi\rangle\simeq 0$.  Subplot (d) displays the local density $\langle\hat{n}_{i,j}\rangle$ as a function of the site numbers along $z$ for the lattice site 20 along $x$. The blue squares (red circles) correspond to the parameters of panel (b) (panel (c), respectively).} \label{Fig:Measure1}
\end{figure}

Since a finite intracavity field is associated with certain atomic density distributions, and vice-versa such distributions are due to the backaction of the cavity field, the signal at the cavity output contains the information on the quantum ground state of the system and permits to monitoring its properties. This situation is quite different from the one encountered when light is used to measure the state of a quantum gas \cite{Rist:2011,Douglas:2012}: There, the mechanical effects of the scattered photons heat the atom and significantly perturb the state. In our case, instead, the mechanical effects of the cavity photons trap the atoms in the BG phase. Cavity losses do not significantly perturb the quantum state of the atoms for the parameter regimes we choose, in which photon number fluctuations can be neglected. We now analyze the signal at the cavity output as a function of the tunneling coefficient at a fixed, fractional density. The corresponding intensity is evaluated by calculating $n_{\rm out}=\langle \hat a_{\rm out}^\dag \hat a_{\rm out}\rangle$, where $\hat a_{\rm out}$ is given in Eq. \eqref{a:out}. The intensity as a function of the tunneling coefficient $t$ is reported in Fig.~\ref{Fig:Measure1}(a):  By increasing the trapping potential depth $V_0$ (decreasing the tunneling) a sudden increase  of the intracavity photon number is observed. This corresponds to the transition to density distributions according to checkerboard clusters, as the subplots (b), (c) and (d) show in detail. Before this sudden increase the density distribution is almost flat along the cavity axis: the atoms delocalize over the lattice sites and there is no coherent scattering of photons into the resonator.

\section{Experimental Parameters}
\label{Sec:Validity}

The Bose-Hubbard Hamiltonian in Eq. \eqref{Hamil_BH} has been derived by performing a series of approximations which have been discussed in detail in the previous section. In this section we show that existing experimental setups, like the one of Ref.~\cite{Ritter,Baumann2010} can observe the phases predicted by Eq. \eqref{Hamil_BH}. Moreover, we identify here the parameters which are then used in the numerical plots presented in Sec. \ref{sec:results}. 

The parameters for the cavity field, which determine the coefficients of the Bose-Hubbard Hamiltonian in Eq. \eqref{Hamil_BH}, are extracted from the experimental values $g_0/2\pi=14.1$ MHz, $\kappa/2\pi=1.3$ MHz, and  $\gamma/2\pi=3$ MHz for $^{87}$Rb atoms \cite{Baumann2010,Ritter}.  The detuning between atoms and the cavity mode at wavelength $\lambda=780$ nm is about $\Delta_a/2\pi=58$ GHz. For these parameters $U_0/\pi\approx 3.4$~kHz. The corresponding value of $S_0$ depends on the Rabi frequency of the transverse laser. For instance, for $\Omega/2\pi=3.08$ MHz then $S_0/2\pi=0.74$ kHz. An external optical lattice trapping the atoms such that the ratio $d_0/\lambda\simeq83/157$ is realized can be made by pumping a cavity mode at wave length $737.7$ nm. Other ratios can be as well considered, depending on the cavity setup and the atoms. 

{\it Parameters.} We now check that these values are consistent with the approximations we made in deriving Eq. \eqref{Hamil_BH}. For this purpose we must fix the number of sites, and thus the number of atoms $N$, since the total shift and the total scattering amplitude must be properly rescaled by $N$. Be $K$ the number of sites, such that for densities $\bar n=1$ the number of sites is equal to the number of atoms. For a one dimensional lattice with $K\simeq 300$ sites one finds $u_0/2\pi=U_0K/2\pi\approx 1.02$ MHz and $s_0/2\pi=S_0\sqrt{K}/2\pi\approx0.013$ kHz, or alternatively $u_0\simeq 0.8\kappa$ and $s_0=0.01\kappa$. Other values are obtained by accordingly changing the Rabi frequency $\Omega$. We set $|\delta_c|=5\kappa$ and observe that for this value $|\delta_c-u_0|\approx |\delta_c|$. We shall now check the order of magnitude of the coefficients of the Bose-Hubbard model for these parameters. Here, $s_0^2K|\delta_c|/(\delta_c^2+\kappa^2)\simeq 0.0045\kappa\simeq 2\pi\times 5.75$kHz. For these parameters the onsite energy due to the cavity field exceeds the MI gap when $\langle\hat\Phi^2\rangle \ge 10^{-3}$. For a two-dimensional lattice with $K=300\times 300$ sites then $u_0=U_0K\simeq 2\pi\times 308.5$ MHz or alternatively $u_0=237\kappa$. For $\Omega/2\pi=2.6$ MHz, for instance, then  $s_0=S_0\sqrt{K}\simeq 0.15\kappa$ and $V_1=\Omega^2/\Delta_a\simeq 0.78$ kHz. For these parameters, typical values of the density distribution give $|\delta_{\rm eff}|\simeq 88 \kappa\gg\kappa$, such that $s_0^2K\delta_{\rm eff}/(\delta_{\rm eff}^2+\kappa^2)=23\kappa$. Here, already for $\langle \hat\Phi^2\rangle\ge 10^{-5}$ cavity backaction has a significant effect. 

{\it Spontaneous emission rate.} Both in the 1D and 2D case, the parameters give a very small occupation of the excited state: The probability that an atom is excited scales with $P_{\rm exc}\sim K {\rm max}(g_0^2n_{\rm cav},\Omega^2)/\Delta_a^2$,  where $n_{\rm cav}$ is the mean intracavity photon number. For the considered parameters $P_{\rm exc}\lesssim 10^{-3}\ll 1$. The corresponding spontaneous emission rate following an excitation due to the cavity field reads $\gamma'_c= \gamma g_0^2 n_{\rm cav}/\Delta_a^2\simeq 2\pi\times 0.17 n_{\rm cav}$ Hz, while spontaneous decay ater an excitation due to the transverse laser scales with $\gamma_L'= \gamma \Omega^2/\Delta_a^2\simeq 2\pi\times 0.08$ Hz. 

{\it Adiabatic elimination of the cavity mode.} We now check the conditions for the adiabatic elimination for the cavity mode for a 1D lattice with 300 sites. The adiabatic elimination of the cavity field from the atomic equations of motion requires that one neglects the coupling with the atoms over the time scale over which the cavity reaches a ``stationary'' value which depends on the instantaneous density distribution. This introduces a time scale $\Delta t=1/|\delta_c+{\rm i}\kappa|$, for which the inequalities shall be fulfilled $S_0\sqrt{n_{\rm cav}}\Delta t\ll 1$ and $U_0n_{\rm cav}\Delta t \ll 1$. These relations are satisfied for the typical numbers of intracavity photons we encounter. In addition, since the atoms must move slowly over this time scale, their kinetic energy (temperature) must be such that $k_BT\ll \hbar/\Delta t$, where $k_B$ is the Boltzmann constant. This latter condition is satisfied for atoms at $T\simeq 1\mu$K, which is achieved in Bose-Einstein condensates.  
 
{\it Neglecting quantum noise.} Quantum noise in Eq. \eqref{a_mf} can be neglected when $Ks_0^2\langle \hat\Phi^2\rangle\gg \kappa^2$, which corresponds to a depth of the lattice created by photon scattering which is much larger than single photon fluctuations. For the parameters here discussed one needs a lattice with sites $K\gg 10^4$, which corresponds to the 2D situation we analyse. The one-dimensional lattice we numerically consider contains $K\simeq 100$ sites, however the scaling of the behaviour with the number of particles show that our predictions remain valid for larger numbers, where one can discard fluctuations in the intracavity photon number.

\begin{figure}
\centering
\includegraphics[width=8cm]{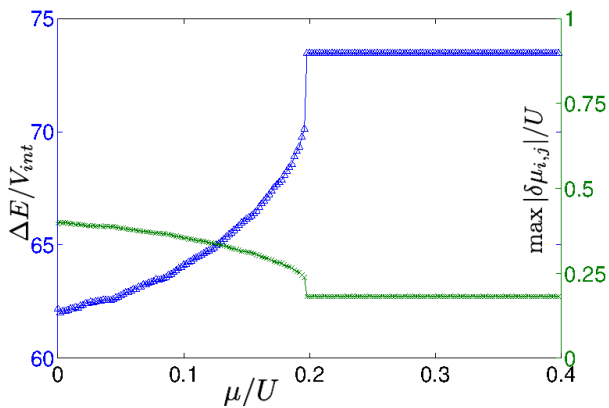}
\caption[]{(color online) Ratio between the energy gap between the two Bloch bands, $\Delta E=\sqrt{4E_R|V_0|}$, and the interaction energy, $V_{\rm int}=U+\max{|\delta\epsilon_{i,j}|}$, as a function of the chemical potential $\mu$ (in units of $U$)and at zero tunneling. The single-band approximation is valid $\Delta E/V_{int}\gg1$. The green curve with crosses shows the maximum values of $\langle\delta\hat\epsilon_{i,j}\rangle$ at the corresponding values of $\mu/U$. The parameters are as same as in Fig.~\ref{Fig:PhaseDiag1}(a).} \label{Fig:validity}
\end{figure}

{\it Single-band approximation.} In the derivation of the Bose-Hubbard Hamiltonian in Eq. \eqref{Hamil_BH} we have performed an expansion of the field operator (\ref{Psi_expand}) into Wannier functions of the lowest band of the external lattice. Discarding the higher bands is correct as the long as the energy gap between a lowest and a first excited Bloch band $\Delta E=\sqrt{4E_R |V_0|}$ is much larger than the interaction energy $V_{\rm int}$, which is here $V_{\rm int}=U+\max|\delta\epsilon_{i,j}|$ between the particles \cite{Jaksch}, where $E_R=\hbar^2 k_0^2/2m$ is the recoil energy. Figure~\ref{Fig:validity} displays the ratio $\Delta E/V_{\rm int}$ in the limit of zero tunneling $t\to0$. We have checked that this ratio remains smaller than unity for the parameters chosen here. Increasing the laser amplitude $\Omega$, i.e., increasing  $s_0$ (and hence $\delta\hat\epsilon$) leads to an increase of $V_{\rm int}$ and thus forces one to take into account higher Bloch bands. %

{\it External harmonic trap.} In our treatment we have assumed the atoms are confined by an external optical lattice. In several experimental situations \cite{Ritter}, however, the atoms are additionally confined by a harmonic trap potential. This gives rise to a smooth position-dependent on-site energy, which will act as biase-field, removing the degeneracy between the two ground-state configurations we identify. Moreover, it can lead to the observation of the characteristic wedding-cake form \cite{Jaksch:BH,WeddingCake}. The influence of cavity backaction on such structures shall be analysed in future works.

\section{Conclusion and outlook}\label{Sec:Conclusion}

Ultracold atoms confined in tight classical lattices and strongly coupled with a standing-wave cavity mode selforganize in order to maximize the number of intracavity photons. This selforganization takes place when the atoms are driven by a transverse laser field which is quasi-resonant with the cavity mode and whose intensity exceeds a threshold value. In this paper we have studied the quantum ground state when the cavity mode has wave length which is incommensurate with the interparticle distance $d_0$ due to the external lattice. We have shown that the atomic density rearranges in clusters, within which the atoms form density waves then locally maximize scattering into the cavity mode. The clusters have mean size corresponding with the beating wave length between the two overlapping fields and are phase locked with one another, so that the intracavity field is maximum. These quantum phases are often characterized by vanishing order parameter and finite compressibility, so that they share several analogies with a Bose-glass phase. 

In our theoretical model, the atomic dynamics are described by a Bose-Hubbard type Hamiltonian, where the effect of the cavity field enters by means of a non-local term, which depends on the density at all sites. This term is the cavity-mediated potential, which depends on the atomic distribution. In particular, its sign is determined by the detunings between atoms and fields, which thus controls whether self-organized structures are energetically favourable. When the sign of the detuning is appropriately chosen, the cavity field gives rise to a long-range interaction between the atoms and to new phases, where the atomic density selforganize in order to maximize the intracavity photon number. The analysis of the type of phase transition leading to these phases requires further careful studies of the system. We finally remark that this system constitutes a novel setting where quantum fluctuations give rise to effects usually associated with disorder.

\section*{Acknowledgments}

The authors are grateful to G. De Chiara, T. Donner, S. Fern\'{a}ndez-Vidal, J. Larson, M. Lewenstein, A. Niederle, and H. Ritsch, for stimulating discussions and helpful comments. This work was supported by the European Commission (IP AQUTE), the European Regional Development Fund, the Spanish Ministerio de Ciencia y Innovaci\'on (QOIT: Consolider-Ingenio 2010; QNLP: FIS2007-66944; FPI; FIS2008-01236; Juan de la Cierva), the Generalitat de Catalunya (SGR2009:00343), the Brazilian Ministry of Science and Technology (MCT), the Conselho Nacional de Desenvolvimento Cientifico e Tecnologico (CNPq), and the German Research Foundation.

\appendix

\section{Derivation of the Bose-Hubbard Hamiltonian}\label{AppBH}

In this appendix we report the basic steps which lead to the derivation of the Bose-Hubbard Hamiltonian in Eq. \eqref{Hamil_BH}. The steps follow methods we developed in Refs. \cite{Larson2008,Fernandez2010}
We first substitute the cavity field operator (\ref{a_mf}), after neglecting the quantum noise term, into the equation for the quantum field operator in Eq. \eqref{Psi1_with_a}. Using the Wannier decomposition, we obtain the equations of motion for operators $\hat b_{l,m}$, that read
\begin{equation}\label{b_l_general}
\dot{\hat b}_{l,m}=\frac{1}{{\rm i}\hbar}[\hat b_{l,m},\hat{\mathcal H}_0+\hat{\mathcal H}_p]-{\rm i}{\mathcal {\hat C}_{l,m}} \,,
\end{equation}
where for $\beta=0,1$
\begin{align}
\hat{\mathcal H}_0=\frac{U}{2}\sum_{i,j}\hat{n}_{i,j}(\hat{n}_{i,j}-1)+(E_0+V_{0}X_0)\hat{N}+(E_1+V_{0}X_1)\hat{B}\,,
\end{align}
is the Bose-Hubbard Hamiltonian in the absence of the cavity field and of the transverse laser, with $E_s$ and $X_s$ given in Eqs. \eqref{Xst}, while $U$ is defined in Eq. \eqref{U}.  Here, $\hat{B}=\hat{B}^x+\hat{B}^z$ is the hopping term, where $\hat{B}^x=\sum_{i,j}(\hat{b}^\dag_{i+1,j}\hat{b}_{i,j}+\hat{b}^\dag_{i,j}\hat{b}_{i+1,j})$ describes tunneling between neighbouring sites of the lattice along $x$ and $\hat{B}^z=\sum_{i,j}(\hat{b}^\dag_{i,j+1}\hat{b}_{i,j}+\hat{b}^\dag_{i,j}\hat{b}_{i,j+1})$ describes tunneling between neighbouring sites of the lattice along $z$.
Hamiltonian $\hat{\mathcal H}_p$ contains the terms due to the pumping laser propagating along the $x$ direction and reads
\begin{align}
\hat{\mathcal H}_p&=V_1\sum_{i,j} J_{0}^{(i,j)} \;\hat{n}_{i,j}+V_1\sum_{i,j} J_{1}^{(i,j)}\hat{B}^x_{i,j} \,,
\end{align}
with $V_1=\hbar\Omega^2/\Delta_a$, while $ J_{0}^{(i,j)}$ is defined in Eq. \eqref{paramwann1} and 
\begin{align}%\label{paramwann1}
J_{1}^{(i,j)}=\int d^2{\bf r}\, w_{i,j}({\bf r})\cos^2(k \,x)\, w_{i+1,j}({\bf r})\nn\,.
\end{align} 
Finally, operator ${\mathcal {\hat C}_{l,m}}$ in Eq. \eqref{b_l_general} is due to the coupling with the cavity field and reads
\begin{align}\label{Cs}
{\mathcal {\hat C}_{l,m}}&=S_0\Big( \frac{S_0\hat{\mathcal Z}}{\hat D^\dag}{\mathcal{\hat P}_{l,m}}+{\mathcal {\hat P}_{l,m}}  \frac{S_0\hat{\mathcal Z}}{\hat D}\Big)+U_0\Big( S_0^2\frac{\hat{\mathcal Z}}{\hat D^\dag}{\mathcal {\hat Q}_{l,m}}\frac{\hat{\mathcal Z}}{\hat D}\Big) \,,
\end{align}
where we have introduced operators $\hat D=(\delta_c-U_0\hat{\mathcal Y})+{\rm i}\kappa$ and
\begin{eqnarray}
{\mathcal {\hat P}_{l,m}}&=&[\hat b_{l,m},\hat{\mathcal{Z}}]\nn\\
 &\approx& Z_{0,0}^{(l,m)}\hat b_{l,m}+Z_{0,1}^{(l,m)}\hat b_{l,m+1}+Z_{1,0}^{(l,m)}\hat b_{l+1,m}\nn\\
 &&+Z_{1,0}^{(l-1,m)}\hat b_{l-1,m}+Z_{0,1}^{(l,m-1)}\hat b_{l,m-1}\,,\nn\\
{\mathcal {\hat Q}_{l,m}}&=&[\hat b_{l,m},\hat{\mathcal{Y}}]\nn\\
 &\approx& Y_{0}^{(l,m)}\hat b_{l,m}+Y_{1}^{(l,m)}\hat b_{l,m+1}+Y_{1}^{(l,m-1)}\hat b_{l,m-1}\,.\nn\\
\end{eqnarray}
Operator ${\mathcal {\hat C}_{l,m}}$ cannot be generally written in the form of the commutator between $\hat b_{l,m}$ and a Hermitian operator.  However, in the thermodynamic limit we have chosen, operator \eqref{Cs} can be cast in the form ${\mathcal {\hat C}_{l,m}}=[\hat{b}_{l,m},\hat{\mathcal H}_{\rm BH}^{(1)}]$ (see for details \cite{Larson2008}), where the operator
\begin{equation}\label{Hamil_BH_1}
\hat{\mathcal H}_{\rm BH}^{(1)}=\sum_{i,j}\Big(\delta\hat{\epsilon}_{i,j} \hat{n}_{i,j}+ \delta \hat{t}_{i,j}^{x} \hat{B}^x_{i,j}+ \delta \hat{t}_{i,j}^{z} \hat{B}^z_{i,j} \Big)
\end{equation}
is different from zero when the pump laser is on, $\Omega>0$. Due to the incommensurate wavelength of laser and cavity mode with respect to the lattice spacing, the coefficients of the Hamiltonian are site-dependent and read
\begin{align}\label{mui}
\delta\hat{\epsilon}_{i,j}=&V_1J_{0}^{(i,j)}+
\frac{\hbar s_0^2}{\hat\delta_{\rm eff}^2+\kappa^2}\hat\Phi\hat\delta_{\rm eff} Z_{0,0}^{(i,j)}\,.
\end{align}
%{\bf GM: changed the onsite energy}
Here, the collective operator $\hat\Phi=1/K\sum_{i,j} Z_{0,0}^{(i,j)}\hat{n}_{i,j}$ appears in the site-dependent parameters. The site-dependent tunneling terms read
\begin{align}\label{t_i}
\delta \hat{t}_{i,j}^{x}=&\, -2\hbar\frac{s_0^2\hat\delta_{\rm eff}}{\hat\delta_{\rm eff}^2+\kappa^2}\hat\Phi  Z_{1,0}^{(i,j)}-V_1 J_{1}^{(i,j)},\nn\\
\delta \hat{t}_{i,j}^{z}=&\,-2\frac{\hbar s_0^2}{\hat\delta_{\rm eff}^2+\kappa^2}\hat\Phi\hat\delta_{\rm eff} Z_{0,1}^{(i,j)}\,.
\end{align}
%{\bf GM: changed the tunneling coefficients}
In the regime of the parameters we consider (see Sec.~\ref{Sec:Validity}) for which $\max|\langle\delta\hat\epsilon_{i,j}\rangle|\sim U$ and $|V_0|\gg\max|\langle\delta\hat\epsilon_{i,j}\rangle|$ (hence the validity of a single-band approximation) $|\delta t^x|$ and $|\delta t^z|$ are at least 8 order of magnitude smaller than $t^{(0)}$ and will be therefore neglected. Thus in our model only the coefficient $\hat\epsilon_{i,j}$ can depend on the lattice sites in a relevant way.

\section{Details on the QMC calculation}\label{AppQMC}
%\section{Quantum Monte Carlo simulation}
For the convenience of the reader, we recapitulate first the basics of the discrete imaginary time world-line algorithm \cite{Batrouni:2} which we are using to obtain the one-dimensional results. Afterwards, we describe how we treat the particular long-range term of our model numerically.

\begin{figure}
\includegraphics[width=0.35\textwidth]{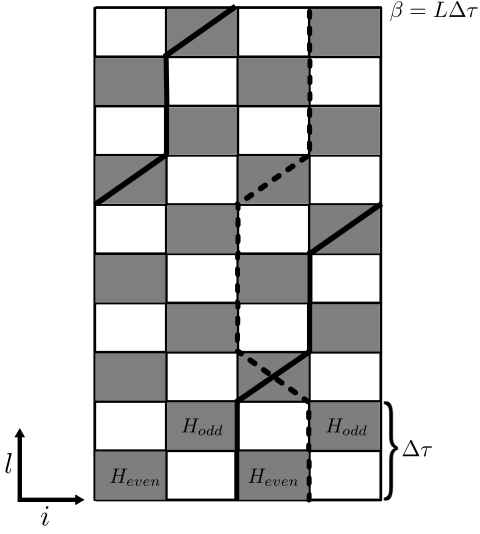}
\caption{\label{fig:QMC} Path integral representation of the system. The horizontal expansion displays the spatial dimension whereas the vertical slices represents the evolution in imaginary time. One valid configuration is described by a set of particle paths (world-lines) that obey periodic boundary conditions in spatial and temporal direction and particle conservation in each slice. World-lines can either exhibit a closed connection from bottom to top (dash line) with so-called winding number \(W=0\) or winded around the system (here with \(W=\!+1\) (solid line)). The shaded squares of the checkerboard pattern indicates the plaquettes, where \(\mathcal{H}_{i,i+1}\) acts and hence hopping can occur.}
\end{figure}

In order to estimate the equilibrium properties of the system, the partition function is decomposed at first by
\begin{equation}
  Z = \mathrm{Tr}\; e^{- \beta \mathcal{H}_{\mathrm{BH}} } =  \mathrm{Tr} \;\left( e^{- \Delta \tau \mathcal{H}_{\mathrm{BH}}} \right)^{L}\,,
\end{equation}
where $\Delta \tau =\beta/L$. In the 1D case, the Bose Hubbard Hamiltonian can be written as a sum of pair Hamiltonians \( \mathcal{H}_{\mathrm{BH}}^{1d} = \sum_{i=0}^{K-1}\mathcal{H}_{i,i+1}\) with:
\begin{align}
%\mathcal{H}_{i,i+1} &= \frac{1}{4}U \left[\hat{n}_{i}(\hat{n}_{i}+1) + \hat{n}_{i+1}(\hat{n}_{i+1}+1) \right] \nonumber\\
%	&+t\left(\hat{a}_i \hat{a}_{i+1} + \mathrm{H.c.}\right) + \frac{1}{2} \mu_{i}\left(\hat{n}_{i} + \hat{n}_{i+1} \right)
\mathcal{H}_{i,i+1} &= \frac{1}{4}U \left[\hat{n}_{i}(\hat{n}_{i}+1) + \hat{n}_{i+1}(\hat{n}_{i+1}+1) \right] \nonumber\\
	&-t\left(\hat{b}^{\dagger}_i \hat{b}_{i+1} + \hat{b}^{\dagger}_{i+1} \hat{b}_{i}\right)\;.% + \frac{1}{2} \left(\delta\epsilon_{i}\hat{n}_{i} + \delta\epsilon_{i+1}\hat{n}_{i+1} \right)\;.
\end{align}
As the second step, the Hamiltonian is typically divided into a part of even site labels, which interact with their subsequent odd labeled sites and a complementary odd part which contains the remaining interactions (\(\mathcal{H}_{\mathrm{BH}}^{1d} = \mathcal{H}_{\mathrm{even}} +  \mathcal{H}_{\mathrm{odd}}\)):
\begin{align}
 \mathcal{H}_{\mathrm{even}} &= \sum_{i=0}^{K/2-1} \mathcal{H}_{2i,2i+1}\\
\mathcal{H}_{\mathrm{odd}} &= \sum_{i=0}^{K/2-1} \mathcal{H}_{2i+1,2i+2}
\end{align}
At this stage, the Suzuki-Trotter \cite{Trotter,Suzuki} formula can be applied:
\begin{equation}
  Z \approx \mathrm{Tr} \left( e^{- \Delta \tau \mathcal{H}_{\mathrm{even}} }  e^{- \Delta \tau \mathcal{H}_{\mathrm{odd}} }\right)^{L}\;,
\end{equation}
which involves an error of the order of \(\mathcal{O}(\Delta \tau^3)\) \cite{DeRaedt}. Finally, \(2L\) sets of occupation number states are inserted between each exponential leading to the effective two-dimensional imaginary time path integral representation of the system:
\begin{equation}
 Z \approx  \sum_{\{n^{l}\}} \prod_{l=0}^{L-1} \langle {n^{2l+2}} | e^{- \Delta \tau \mathcal{H}_{\mathrm{even}} } | {n^{2l+1}} \rangle \langle {n^{2l+1}} |  e^{- \Delta \tau \mathcal{H}_{\mathrm{odd}} } | {n^{2l}} \rangle,
\end{equation}
in which the bosons are described by world-lines (Fig.~\ref{fig:QMC}). The system is now described by \(2L\) matrix elements, that consists of easy calculable two-site problems. The Monte Carlo process enters now by sampling the world-lines via local updates, where the thermodynamical weights of only four plaquettes are involved and optionally global updates, that are needed to insert and delete straight world-lines in the system for the grand-canonical simulation \cite{Batrouni:1, Batrouni:2}.

The quantities, that are diagonal in the occupation-number representation, like the density (\ref{eq:observable2}) and the local density fluctuation (\ref{eq:observable3}) can now estimated easily by averaging the occupation numbers \(n(i,l)\) over all slices after reaching the equilibrium:
\begin{align}
	\label{eq:observable1}
	\langle \hat{n}_i \rangle &= \frac{1}{2L}\sum_{l=0}^{2L-1} \langle n(i,l) \rangle\\
	\label{eq:observable2}
	\bar{n} &= \frac{1}{K} \sum_{i=0}^{K-1} \langle \hat{n}_i \rangle\\
	\label{eq:observable3}
	\kappa_{i} &= \langle \hat{n}_i^2 \rangle - \langle \hat{n}_i \rangle^2\;,
\end{align}
where \(\langle \dots \rangle\) denotes the Monte Carlo average.
Characterizing the superfluidity is a bit more cumbersome.
It has been discussed in Refs.~\cite{PollockCeperley, Batrouni:2} that the SF density is related to the mean-square of the winding number~\(\langle W^2 \rangle\) by:
\begin{equation}
  \label{eq:SFdens}
  \rho_s = \frac{\langle W^2 \rangle}{2\beta\, t }K\;,
\end{equation}
whereas the winding number is defined as the number of world-lines that are winded along the lattice due to the periodic boundary conditions (see Fig. \ref{fig:QMC}). In order to determine \(\rho_s\), one commonly defines the pseudo-current, that counts the number of particles that moves to the right minus the number of particles moving to the left for every time slice $l=[0,2L-1]$:
\begin{align}
 j(l)  = \frac{1}{2} \sum_{i=0}^{K-1} & \left[ n(i,l) - n(i+1,l) \right] \nn\\
&-\left[n(i,l+1)- n(i+1,l+1)\right] \,.
\end{align}
The mean-square winding number can now computed via:
\(
  \langle W^2\rangle
  %=  \langle \left(\frac{1}{K} \sum_{l=0}^{2L-1} j(l) \right)^2\rangle
  =  \frac{1}{K^2} \sum_{l,l'=0}^{2L-1} \langle j(l)j(l') \rangle\,.
\)
Due to the fact, that the applied (local and global) Monte Carlo updates conserve the winding number in every step, \(W\) is technically restricted to zero. 
In order to determine \(\langle W^2 \rangle\) anyway, one computes the correlator  \( J(l) = \langle j(l) j(0)\rangle\) during the simulation and its Fourier transform
\(
  \tilde{J}(\omega_n) = \sum_l e^{i\,l\omega_n} J(l)
\)
afterwards. The value at zero obeys
\begin{equation}
\label{eq:Jtozero}
  \tilde{J}(0) = \langle \sum_l  j(l) j(0) \rangle =\frac{K^2}{2L} \langle W^2 \rangle\;,
\end{equation}
that enables the determination of the SF density by means of Eq.(\ref{eq:Jtozero}), (\ref{eq:SFdens}) and \(\beta = \Delta \tau L\) via the extrapolation:
\begin{equation}
\rho_s = \lim_{\omega_n \rightarrow 0}  \frac{\tilde{J}(\omega_n)}{t \Delta\tau  K}\;.
\end{equation}

In our particular model (\ref{Hamil_BH:1D}), the long-range interaction term:
\begin{equation}
\hat{V} = \frac{ \hbar s_0^2}{\hat\delta_{\rm eff}^2+\kappa^2} K\hat\Phi^2\hat\delta_{\rm eff}
\end{equation}
occurs and has to be treated additionally. We decided to distribute this term equally to both the even and the odd part of the Hamiltonian, namely \(\mathcal{H}_{s} = \hat{U}_{s} + \hat{T}_{s} + \hat{V}/2\), where \(s=\{\mathrm{even},\mathrm{odd}\}\). The (symmetric) Trotter decomposition of the matrix element of slice \(l\) is then given by:
\begin{equation}
\label{eq:matrixelement_lr}
\langle n^{l+1} | e^{- \Delta \tau \mathcal{H}_{s} } | n^l \rangle \approx \langle n^{l+1} | e^{- \frac{\Delta \tau}{4} \hat{V}}  e^{-  \frac{\Delta \tau}{2} \hat{U}_{s}}(1- \Delta \tau \hat{T}_{s} ) e^{-  \frac{\Delta \tau}{2} \hat{U}_{s}}  e^{- \frac{\Delta \tau}{4}\hat{V}} | n^l \rangle\;,
\end{equation}
Since \(\hat{V}\) is diagonal in the occupation-number representation, one can trace back the matrix element (\ref{eq:matrixelement_lr}) to the common case of the short-range Bose-Hubbard model:
\begin{align}
&\langle n^{l+1} | e^{- \Delta \tau \mathcal{H}_{s} } | n^l \rangle \nonumber\\
&\quad \approx e^{-\frac{\Delta \tau}{4} (V(\{n^{l+1}\}) + V(\{n^{l}\}))} \prod_{i\; \mathrm{even} \atop (i\; \mathrm{odd})} \langle n^{l+1}_i n^{l+1}_{i+1} | \mathcal{H}_{i,i+1} | n^l_i n^l_{i+1} \rangle
\end{align}
Within the Monte Carlo sampling, ratios of these the matrix elements has to evaluated efficiently. In the present model, one make use of the fact, that the interaction is a function only two accumulative quantities, which depends linearly on the occupation numbers:
\begin{equation}
%\langle n| \hat{A}(\{\hat{n}_i\}) | n \rangle = f\left(\sum_{i,j} Z_0^{(i)} Z_0^{(j)} n_i n_j,  \sum_{i,j} Y_0^{(i)} Y_0^{(j)} n_i n_j \right)\;,
\langle n| \hat{V} | n \rangle = V(\{n\})=f\left(\frac{1}{K}\sum_{i} Z_0^{(i)}  n_i ,  \frac{1}{K} \sum_{i} Y_0^{(i)} n_i \right)\;,
\end{equation}
where the (non-linear) function is:
\begin{equation}
f(\Phi,\Psi) = \frac{\hbar s_0^2}{(\delta_c - u_0\Psi)^2 + \kappa^2}K \Phi^2 (\delta_c - u_0\Psi)\;.
\end{equation}
For the case of single particle updates, each of these quantities can evaluated in constant computer time, if one uses the auxillary variables \(\Phi = \frac{1}{K} \sum_i  Z_0^{(i)} n_i \) and  \(\Psi = \frac{1}{K} \sum_i  Y_0^{(i)} n_i \) and only tracks the differences caused by the update. For an update, e.g. \(|n_l,n_{l+1} \rangle \rightarrow |n'_l,n'_{l+1} \rangle = |n_l +1 ,n_{l+1} - 1\rangle \), one has to compute:
\begin{align}
\Phi' &= \Phi + (Z_0^{(l)}-  Z_0^{(l+1)})/K\\
\Psi' &= \Psi + (Y_0^{(l)}  - Y_0^{(l+1)})/K\;.
\end{align}
The desired matrix element \(\langle n'| \hat{V} | n' \rangle\) can now calculated by applying the new auxillary variable in the function \(f(\Phi', \Psi')\). The auxillary variables has to allocated for every time slice separately.


\begin{thebibliography}{breitestes Label}

\bibitem{Review_RMP}
%Cold atoms in cavity-generated dynamical optical potentials
H. Ritsch, P. Domokos, F. Brennecke, and T. Esslinger, 
Rev. Mod. Phys. {\bf 85}, 553 (2013).

\bibitem{Deutsch:1995}
I. H. Deutsch, R. J. C. Spreeuw, S. L. Rolston, and W. D. Phillips, Phys. Rev. A {\bf 52}, 1394 (1995).

\bibitem{Phillips}
%Bragg Scattering from Atoms in Optical Lattices
G. Birkl, M. Gatzke, I. H. Deutsch, S. L. Rolston, and W. D. Phillips, Phys. Rev. Lett. {\bf 75}, 2823 (1995). 

\bibitem{Hemmerich}
%Bragg Diffraction in an Atomic Lattice Bound by Light
M. Weidem\"uller, A. Hemmerich, A. G\"orlitz, T. Esslinger, and T. W. H\"ansch,
Phys. Rev. Lett. {\bf 75}, 4583 (1995); 	
%Local and global properties of light-bound atomic lattices investigated by Bragg diffraction
M. Weidem\"uller, A. G\"orlitz, T. W. H\"ansch, and A. Hemmerich, Phys. Rev. A {\bf 58}, 4647 (1998).

\bibitem{Domokos2002}
%Collective Cooling and Self-Organization of Atoms in a Cavity
P. Domokos, and H. Ritsch, Phys. Rev. Lett. \textbf{89}, 253003 (2002).

\bibitem{Black2003}
%Observation of Collective Friction Forces due to Spatial Self-Organization of Atoms: From Rayleigh to Bragg Scattering
A.T. Black, H.W. Chan, and V. Vuleti\'{c}, Phys. Rev. Lett. \textbf{91}, 203001 (2003).

\bibitem{Baumann2010}
%The Dicke Quantum Phase Transition with a Superfluid Gas in an Optical Cavity
K. Baumann, C. Guerlin, F. Brennecke, and T. Esslinger, Nature {\bf 464}, 1301 (2010).

\bibitem{Barrett2012}
%Self-Organization Threshold Scaling for Thermal Atoms Coupled to a Cavity
K. J. Arnold, M. P. Baden, and M. D. Barrett,
Phys. Rev. Lett. {\bf 109}, 153002 (2012).

\bibitem{CARL:1}
 R. Bonifacio and L. de Salvo, Nucl. Instrum. Methods Phys. Res., Sect. A {\bf 341}, 360 (1994); R. Bonifacio, L. De Salvo, L. M. Narducci, and E. J. D'Angelo, Phys. Rev. A {\bf 50}, 1716 (1994).
 
 \bibitem{CARL:2}
 D. Kruse, C. von Cube, C. Zimmermann, and Ph. W. Courteille, Phys. Rev. Lett. {\bf 91}, 183601 (2003)
 

\bibitem{Kuramoto}
%Self-Synchronization and Dissipation-Induced Threshold in Collective Atomic Recoil Lasing
C. von Cube, S. Slama, D. Kruse, C. Zimmermann, Ph. W. Courteille, G. R. M. Robb, N. Piovella, and and R. Bonifacio, Phys. Rev. Lett. {\bf 93}, 083601 (2004).

\bibitem{Hemmerich:Bistable}
B. Nagorny, Th. Els\"asser, and A. Hemmerich, Phys. Rev. Lett. {\bf 91}, 153003 (2003); Th. Els\"asser, B. Nagorny, and A. Hemmerich, Phys. Rev. A {\bf 69}, 033403 (2004).

\bibitem{Stamper-Kurn}
S. Gupta, K. L. Moore, K. W. Murch, and D. M. Stamper-Kurn, Phys. Rev. Lett. {\bf 99}, 213601 (2007); T.P. Purdy, D.W.C. Brooks, T. Botter, N. Brahms, Z.-Y. Ma, and D.M. Stamper-Kurn, Phys. Rev. Lett. {\bf 105}, 133602 (2010).

\bibitem{Ritter}
F. Brennecke, S. Ritter, T. Donner, and T. Esslinger, Science 322, 235 (2008); S. Ritter, F. Brennecke, K. Baumann, T. Donner, C. Guerlin, and T. Esslinger, Appl. Phys. B 95, 213 (2009).

\bibitem{Larson:2008}
J. Larson, G. Morigi, and M. Lewenstein, Phys. Rev. A {\bf 78}, 023815 (2008).

\bibitem{Lev}
%Emergent crystallinity and frustration with Bose–Einstein condensates in multimode cavities
S. Gopalakrishnan, B. L. Lev, and P. M. Goldbart, Nat. Phys. {\bf 5}, 845 (2009). 
\bibitem{Goldbart}
%Frustration and Glassiness in Spin Models with Cavity-Mediated Interactions
S. Gopalakrishnan, B. L. Lev, and P. M. Goldbart, Phys. Rev. Lett. {\bf 107}, 277201 (2011);
%Dicke Quantum Spin Glass of Atoms and Photons
P. Strack, and S. Sachdev, Phys. Rev. Lett. {\bf 107}, 277202 (2011). 

\bibitem{Hessam:PRL}
%Bose-Glass Phases of Ultracold Atoms due to Cavity Backaction
H. Habibian, A. Winter, S. Paganelli, H. Rieger, and G. Morigi, Phys. Rev. Lett. {\bf 110}, 075304 (2013). 

\bibitem{Habibian2011}
%Quantum light by atomic arrays in optical resonators
H. Habibian, S. Zippilli, and G. Morigi, Phys. Rev. A \textbf{84}, 033829 (2011).

\bibitem{Fisher}
%Boson localization and the superfluid-insulator transition
M.P.A. Fisher, P.B. Weichman, G. Grinstein, and D.S. Fisher, Phys. Rev. B \textbf{40}, 546-570 (1989).

\bibitem{BoseGlass}
%Localization in Interacting, Disordered, Bose Systems
R.T. Scalettar, G.G. Batrouni, and G.T. Zimanyi, Phys. Rev. Lett. {\bf 66}, 3144 (1991); 
%Superfluid-Insulator Transition in Disordered Boson Systems
W. Krauth, N. Trivedi, and D. Ceperley, Phys. Rev. Lett. {\bf 67}, 2307 (1991).

\bibitem{Deissler2010}
%Delocalization of a disordered bosonic system by repulsive interactions
B. Deissler, M. Zaccanti, G. Roati, C. D'Errico, M. Fattori, M. Modugno, G. Modugno, and M. Inguscio, Nat. Phys. {\bf 6}, 354 (2010).

\bibitem{Fernandez2010}
%Quantum ground state of self-organized atomic crystals in optical resonators
S. Fern\'{a}ndez-Vidal, G. De Chiara, J. Larson, and G. Morigi, Phys. Rev. A \textbf{81}, 043407 (2010).

\bibitem{Larson2008}
% %Mott-Insulator States of Ultracold Atoms in Optical Resonators
J. Larson, B. Damski, G. Morigi, and M. Lewenstein, Phys. Rev. Lett. \textbf{100}, 050401 (2008); 
%Quantum stability of insulator-like states of cold atoms in optical resonators
J. Larson, S. Fernandez-Vidal, G. Morigi, and M. Lewenstein,
New Journal of Physics {\bf 10}, 045002 (2008).

\bibitem{HLE}
D.F. Walls and G.J. Milburn, \textit{Quantum Optics} (Springer, Berlin, 1994).

\bibitem{Maschler}
%Cold Atom Dynamics in a Quantum Optical Lattice Potential
C. Maschler, and H. Ritsch, Phys. Rev. Lett. \textbf{95}, 260401 (2005).

\bibitem{MekhovPRL}
%Probing quantum phases of ultracold atoms in optical lattices by transmission spectra in cavity quantum electrodynamics
I. B. Mekhov, C. Maschler, and H. Ritsch, Nat. Physics {\bf 3}, 319 (2007); 
%Cavity-Enhanced Light Scattering in Optical Lattices to Probe Atomic Quantum Statistics
Phys. Rev. Lett. {\bf 98}, 100402 (2007).

\bibitem{Asboth2005}
%Self-organization of atoms in a cavity field: Threshold, bistability, and scaling laws
J. K. Asb\'oth, P. Domokos, H. Ritsch, and A. Vukics, Phys. Rev. A {\bf 72}, 053417 (2005).

\bibitem{Jaksch:BH}
%Cold Bosonic Atoms in Optical Lattices
D. Jaksch, C. Bruder, J. I. Cirac, C. W. Gardiner, and P. Zoller,
Phys. Rev. Lett. {\bf 81}, 3108 (1998).

\bibitem{Bloch:RMP}
%Many-body physics with ultracold gases
I. Bloch, J. Dalibard, and W. Zwerger,
Rev. Mod. Phys. {\bf 80}, 885 (2008).

\bibitem{Homodyne}
%Homodyne Detection and Quantum State Reconstruction
D.-G. Welsch, W. Vogel, and T. Opatrny, Progr. in Opt., Vol. XXXIX, ed. E. Wolf (Elsevier, Amsterdam, 1999), p. 63--211.

\bibitem{Nagy2008}
%Self-organization of a Bose-Einstein condensate in an optical cavity, 
D. Nagy, G. Szirmai, and P. Domokos, Eur. Phys. J. D {\bf 48}, 127 (2008). 

\bibitem{Keeling2010}
%Collective Dynamics of Bose-Einstein Condensates in Optical Cavities
J. Keeling, M. J. Bhaseen, and B. D. Simons, Phys. Rev. Lett. {\bf 105}, 043001 (2010).

\bibitem{Damski}
%Atomic Bose and Anderson Glasses in Optical Lattices
B. Damski, J. Zakrzewski, L. Santos, P. Zoller, and M. Lewenstein, Phys. Rev. Lett. \textbf{91}, 080403 (2003).

\bibitem{Meystre}
%Bistable Mott-insulator to superfluid phase transition in cavity optomechanics
W. Chen, K. Zhang, D. S. Goldbaum, M. Bhattacharya, and P. Meystre, Phys. Rev. A 80, 011801 (2009).

\bibitem{StamperKurn}
%S. Gupta, K. L. Moore, K. W. Murch, and D. M. Stamper-Kurn, Phys. Rev. Lett. {\bf 99}, 213601 (2007); 
K. W. Murch, K. L. Moore, S. Gupta, and D. M. Stamper-Kurn, Nature Phys. {\bf 4}, 561 (2008).

\bibitem{Deng08}
%Phase diagram and momentum distribution of an interacting Bose gas in a bichromatic lattice
X. Deng, R. Citro, A. Minguzzi, E. Orignac, Phys. Rev. A {\bf 78}, 013625 (2008).

\bibitem{Roux08}
%Quasiperiodic Bose-Hubbard model and localization in one-dimensional cold atomic gases
G. Roux, T. Barthel, I.P. McCulloch, C. Kollath, U. Schollw\"{o}ck, T. Giamarchi, Phys. Rev. A {\bf 78}, 023628 (2008).

\bibitem{Volz_03}
%Characterization of elastic scattering near a Feshbach resonance in 87Rb
T. Volz, S. D\"{u}rr, S. Ernst, A. Marte, and G. Rempe, Phys. Rev. A \textbf{68}, 010702(R) (2003).

\bibitem{Kruger_07}
%Critical Point of an Interacting Two-Dimensional Atomic Bose Gas
P. Kr\"{u}ger, Z. Hadzibabic, and J. Dalibard, Phys. Rev. Lett. \textbf{99}, 040402 (2007).

\bibitem{Footnote:QMC} We checked these results by comparing them with the curves obtained performing canonical and grand-canonical Monte Carlo simulations. To prevent that the simulation is being trapped in a meta-stable state, we annealed the temperature in the simulation from higher temperature till $T=0$. We also checked the results for system sizes which are multiples of 74 (this choice respects the underlying quasi-periodicity), and found only small deviations from the results reported in Fig.~\ref{Fig:N_mu_Dc}.  

\bibitem{Mottl}
%Exploring Symmetry Breaking at the Dicke Quantum Phase Transition
K. Baumann, R. Mottl, F. Brennecke, and T. Esslinger, Phys. Rev. Lett. \textbf{107}, 140402 (2011).

\bibitem{Footnote:plateaus}
In Fig. \ref{Fig:Dens_twoConfig}(a) we do not observe the plateaus with fractional filling, which were instead found in Roux {\it et al.} \cite{Roux08}, which is justified since our parameter regime is different from the one considered in  Ref. \cite{Roux08}. We have verified that these plateaus are found in our model by increasing $s_0\to10s_0$ and $U\to100U$.

\bibitem{Footnote:QMC:2}
To analyse these jumps from the incompressible phase to the fractional filling,
we performed a canonical QMC simulation, in which the chemical potential is calculated by the difference of the internal energies $\mu = (E(N + 1) - E(N))K/U$ of two simulations with $N$ and $N+1$ particles. We found regions with negative compressibility (i.e., negative slope in the $\bar{n}-\mu$ plot). We have then evaluated the grand-potential $E-\bar{n}\mu/U$ calculated from the internal energy per site $E$, whose minimum is a criterion for the stability of the system state for temperature $T=0$: We found that the region with negative compressibility of the simulations with the canonical ensemble are associated with metastable branches. In contrast, the grand-canonical simulation does follow the stable branch that results in a kink of the grand-potential, which is typical for a system undergoing a first-order phase transition. 

\bibitem{Batrouni:1}
%Phase transitions in an interacting boson model with near-neighbor repulsion.
P. Niyaz, R.T. Scalettar, C.Y. Fong, G.G. Batrouni, Phys. Rev. B {\bf 50}, 362 (1994).

\bibitem{Batrouni:2}
%Quantum Critical Phenomena in One-Dimensional Bose Systems
G.G. Batrouni, R.T. Scalettar, and G.T. Zimanyi, Phys. Rev. Lett. {\bf 65}, 1765 (1990); 
%World-line quantum Monte Carlo algorithm for a one-dimensional Bose model
G.G. Batrouni, R.T. Scalettar, Phys. Rev. B {\bf 46}, 9051 (1992).

%\bibitem{Batrouni_Maxwell}
%%Phase Separation in Supersolids
%G.G. Batrouni, and R.T. Scalettar, Phys. Rev. Lett. {\bf 84}, 1599 (2000); 
%%Ring exchange and phase separation in the two-dimensional boson Hubbard model
%V.G. Rousseau, R.T. Scalettar, Phys. Rev. B {\bf 72}, 054524 (2005).

\bibitem{Sheshadri}
%Superfluid and Insulating Phases in an Interacting-Boson Model: Mean-Field Theory and the RPA
K. Sheshadri, H. R. Krishnamurthy, R. Pandit, and T. V. Ramakrishnan, Europhys. Lett. \textbf{22}, 257 (1993);
%Spectral weight redistribution in strongly correlated bosons in optical lattices
C. Menotti and N. Trivedi,
Phys. Rev. B {\bf 77}, 235120 (2008). 


\bibitem{Rist:2011}
%Light scattering by ultracold atoms in an optical lattice
S. Rist, C. Menotti, and G. Morigi,
Phys. Rev. A {\bf 81}, 013404 (2010).

\bibitem{Douglas:2012}
%Light scattering from ultracold atomic gases in optical lattices at finite temperature
J. S. Douglas and K. Burnett,
Phys. Rev. A {\bf 84}, 033637 (2011). 

\bibitem{Jaksch}
D. Jaksch, V. Venturi, J.I. Cirac, C.J. Williams, P. Zoller, Phys. Rev. Lett. \textbf{89}, 040402 (2002).

\bibitem{WeddingCake}
M. Lewenstein, A. Sanpera, V. Ahufinger, B. Damski, A. Sen De, and U. Sen,
Adv. Phys. {\bf 56}, 243 (2007).


\bibitem{Trotter}
H. F. Trotter, Proc. Am. Math. Soc. {\bf 10}, 545 (1959).

\bibitem{Suzuki}
%Generalized Trotter's Formula and Systematic Approximants of Exponential Operators and Inner Derivations with Applications to Many-Body Problems
M. Suzuki, Commun. Math. Phys. {\bf 51}, 183 (1976).

\bibitem{DeRaedt}
%Applications of the generalized Trotter formula
H. De Raedt and B.De Raedt, Phys. Rev. A {\bf 28}, 3575 (1983).

\bibitem{PollockCeperley} 
%Path-integral computation of superfluid densities
%SFdensity from <W^2>
E. L. Pollock and D. M. Ceperley, Phys. Rev. B {\bf36}, 8343 (1987).


\end{thebibliography}
\end{document}